\documentclass[aps,prd,nofootinbib,showpacs,showkeys]{revtex4-1}

\usepackage{graphicx}
\usepackage{dcolumn}
\usepackage{bm}
\usepackage{multirow}
\usepackage{color}%
\usepackage[english]{babel}
\usepackage{amsmath,amssymb}
\usepackage{amsfonts}
\usepackage{mathrsfs}
\usepackage{epsfig}
\usepackage{tabularx}
\usepackage{array}
\usepackage{graphics} 
\usepackage{mathtools} 
\usepackage{yhmath} 
\usepackage{tabularx} 

\def\vok{\mathrel{\rlap{\lower0pt\hbox{\hskip1pt{$k$}}}
    \raise6pt\hbox{$\neg$}}}
\def\voc{\mathrel{\rlap{\lower0pt\hbox{\hskip1pt{$c$}}} 
    \raise4pt\hbox{$\neg$}}}

\newcommand{\refe}[1]{Eqn.~(#1)} 
\newcommand{\refs}[1]{Ref.~#1} 
\newcommand{\reft}[1]{Table~#1} 
\newcommand{\refp}[1]{\mbox{Section~#1}} 
\newcommand{\refa}[1]{Appendix~#1} 
\begin{document}


\title{CPT-odd Photon in Vacuum-Orthogonal Model }

\newcommand{\metu}{Department of Physics, Middle East Technical University, Ankara 06800, Turkey.}

\author{ Soner Albayrak}\email{alsoner@metu.edu.tr}  \affiliation{ \metu }
\author{Ismail Turan}\email{ituran@metu.edu.tr}  \affiliation{ \metu }


\date{\today}

\begin{abstract}
Lorentz and CPT violation in the photon sector with the operators of arbitrary mass dimension has been proposed in the context of Standard-Model Extension. The CPT-odd subset of this enlarged model is investigated from a quantum-field theoretical point of view. The generic forms of dispersion relations, polarization vectors and the propagators are obtained for this special subset. Particularly, the general vacuum-orthogonal model is analyzed and it is demonstrated that the vacuum orthogonal model remains vacuum orthogonal at all orders. Although the model has birefringent solutions, they are shown to be spurious. Furthermore, the relevant polarization vectors are shown to be conventional. Leading order model is explicitly analyzed and it is demonstrated that there exists a nontrivial coefficient space satisfying above properties.
\end{abstract}


\maketitle

 
 
\section{Introduction}
\label{intro}
The interest in exotic theories beyond the Standard Model (SM) has been increasing over the past few decades. Among the motivations is the exploration of a phenomena not explainable either by General Relativity or Quantum Field Theory, as that would be a direct probe into the ultimate quantum theory of gravity. One of the leading theoretical candidates for such a phenomena is the violation of Lorentz and CPT symmetries. Indeed, most of the current approaches to the quantum gravity naturally allow the violation of these symmetries, among whom string theories \cite{String_Theory}, loop quantum gravity \cite{Loop QG,Gambini}, noncommutative field theories \cite{Non-commutative G}, and many others \cite{LV Theories} can be named.

The breaking of the Lorentz symmetry may be either exact or spontaneous, although it was shown that the usual Riemann geometry cannot be maintained in the gravity sector when the breaking is explicit \cite{Gravity}\footnote{Explicit Lorentz symmetry breaking might suggest alternative geometries like Riemann-Finsler \cite{Kostelecky:2011qz}.}. As for the spontaneous breaking, various approaches about insertion of the Lorentz violation into the model, among which modifications in transformation laws \cite{Kinematical LV} and field theoretical approaches \cite{Field T. LV,Gambini} can be named, have been pursued in the literature, albeit such different approaches can be shown to be contained in a systematic field theoretical framework \cite{Fermion,Kostelecky2009}.

The systematic framework for exploration of Lorentz and CPT violations was constructed over 15 years ago. This framework, so called \emph{Standard Model Extension} (SME) \cite{SME,Colladay.Kostelecky,Gravity}, is an action level \emph{effective field theoretical} (EFT) approach in which Lorentz violation is inserted to the model via background fields named \emph{Lorentz Violating Terms} (LVT), and has been analyzed and investigated both in theoretical and experimental fronts \cite{Kostelecky:2008ts} and references therein. Basically, it is assumed that the effective low energy description of the high energy fundamental theory can be expanded in energy over a mass scale, which is possibly related to the Planck scale. In this expansion, the lowest order term becomes the Standard Model. With the EFT approach, next terms in this expansion can be examined with the field theoretical machinery built within the SM. The task of examination of the next-to-leading term, called \emph{minimal extension of Standard Model} ({\it m}SME) \cite{Colladay.Kostelecky}, has been undergone in all sectors. While {\it m}SME constitutes all renormalizable operators, as gravity itself is nonrenormalizable, it is reasonable for the next term in the expansion to constitute of nonrenormalizable operators of arbitrarily high mass dimensions, called \emph{non-minimal Standard Model Extension} ({\it nm}SME). The photon, neutrino and the fermions sectors of {\it nm}SME were introduced in 2009, 2011, 2013 respectively \cite{Kostelecky2009,Neutrino,Fermion}. There are data tables \cite{Kostelecky:2008ts} listing all the available bounds on the sectors of  {\it m}SME and {\it nm}SME. The updates of the tables are given in \cite{Kostelecky:2008tsupdates}.

In the past, nonrenormalizable theories have not been considered very popular. This attitude changed as EFT approach to the nonrenormalizable theories has been proven to be quite useful \cite{EFT}. The reasoning beyond the EFT lies within the assumption of small deviations, which actually determines a validity range hence justifies the name ``effective". In the case of {\it nm}SME, the current bounds on the LVT directly indicate the necessity of quite small deviations in the interested low energy regimes, thus suggests the use of nonrenormalizable LVT within EFT.

The available LVT in each sector of the SME splits into two parts: those which violate CPT invariance, being called \emph{CPT-odd}; and those which do not, being called \emph{CPT-even}. Among these sectors, the CPT-even and CPT-odd modified photons have been studied in the {\it m}SME \cite{Kostelecky:2001mb}. The photon sector of the {\it nm}SME has been discussed in \refs{\cite{Kostelecky2009}} and compactly analyzed in \refs{\cite{Cambiaso:2012vb}}. The CPT-even part of it has been specifically studied in \refs{\cite{Shreck2013,Schreck:2013gma}}. A similar discussion for the CPT-odd modified photon part of the {\it nm}SME is missing in literature. Hence, the aim of this study is to fill this gap by doing the analysis of CPT-odd modified photon from a quantum field theoretical point of view.

In the photon sector of the {\it nm}SME, the CPT-odd and CPT-even contributions are denoted by the coefficients  $\hat{k}_{AF}$ and $\hat{k}_{F}$, respectively. The symbol hat ``$\hat{\;\;}$" is used to indicate that all higher order terms are contained. As it provides a natural classification with direct relevance to observations and experiments, the decomposition of these coefficients into spin weighted spherical harmonics, called spherical decomposition, is introduced in \refs{\cite{Kostelecky2009}}. Then, the LVT $\hat{k}_{F}$ and $\hat{k}_{AF}$ decompose as
\begin{equation}
\begin{aligned}
\hat{k}_{AF}&\longrightarrow\left\{(k_{AF}^{(d)})_{njm}^{(0B)}, (k_{AF}^{(d)})_{njm}^{(1B)}, (\vok_{AF}^{(d)})_{njm}^{(1E)}\right\}\,,\\
\hat{k}_F&\longrightarrow\left\{(c_F^{(d)})_{njm}^{(0E)}, (k_F^{(d)})_{njm}^{(0E)}, (\vok_F^{(d)})_{njm}^{(1E)}, (\vok_F^{(d)})_{njm}^{(2E)}, (k_F^{(d)})_{njm}^{(1B)}, (\vok_F^{(d)})_{njm}^{(2B)}\right\}\,,
\end{aligned}\label{Eq:decompose of k}
\end{equation}
where $c$ denotes nonbirefringence, and negation diacritic denotes vacuum-orthogonality (no leading order effect on vacuum propagation). The symbols $n, j, m$ denote frequency dependence, total angular momentum, $z$-component of angular momentum respectively; whereas $E$ and $B$ refer to the parity of the operator, and the preceding number gives the spin weight of the operator. These coefficients can be regrouped according to their effects on the leading order vacuum propagation. That splits the overall coefficient space into two distinct parts as listed in \reft{\ref{Table-sdc}}.
\begin{table*}
\caption{\label{Table-sdc}Spherically decomposed coefficients according to their vacuum properties.}
\begin{ruledtabular}
\begin{tabular}{l@{\hskip 0.3in}c@{\hskip 0.3in}c}
& $\bm{\hat{k}_{AF}}$ & $\bm{\hat{k}_{F}}$\\\hline
Vacuum Models &$k_{(V)jm}^{(d)}$&$c_{(I)jm}^{(d)}$, $k_{(E)jm}^{(d)}$, $k_{(B)jm}^{(d)}$\\\hline
Vacuum-Orthogonal\\ \hskip 1cm Models &$(\vok_{AF}^{(d)})_{njm}^{(0B)}$, $(\vok_{AF}^{(d)})_{njm}^{(1B)}$, $(\vok_{AF}^{(d)})_{njm}^{(1E)}$&$(\voc_{F}^{(d)})_{njm}^{(0E)}$, $(\vok_{F}^{(d)})_{njm}^{(0E)}$, $(\vok_{F}^{(d)})_{njm}^{(1E)}$,\\&& $(\vok_{F}^{(d)})_{njm}^{(2E)}$, $(\vok_{F}^{(d)})_{njm}^{(1B)}$, $(\vok_{F}^{(d)})_{njm}^{(2B)}$
\end{tabular}
\end{ruledtabular}
\end{table*}

The outline of the paper is as follows. In \refp{\ref{sec:COPS}}, the modified dispersion relations is investigated for the generic coefficient subspace. From there, the attention is restricted to a particular subset, and the solutions for the photon field $A^\mu$ and the modified propagator are analyzed. In \refp{\ref{sec:VOMFCOP}}, we further restrict the coefficient space to vacuum-orthogonal LVT only, and prove that vacuum orthogonal model remains vacuum orthogonal at all orders. We demonstrate that the dispersion relations for this models split into two sets, non-conventional and conventional; and non-conventional dispersion relations are shown to be spurious, whereas conventional dispersion relations are shown to accept conventional polarization vectors. In \refp{\ref{sec:SMA}}, we analyze some special cases and show that there exists a nontrivial coefficient subspace satisfying above results.

\section{The CPT-odd Extension of Photon Sector}
\label{sec:COPS}
The general form of {\it nm}SME Lagrangian for the photon sector can be read off from \refe{8} of \refs{\cite{Kostelecky2009}}. For the model with the CPT-odd LVT only ($\hat{k}_{F}=0$), the Lagrangian becomes
\begin{equation}
\mathcal{L}={}-\frac{1}{4}F_{\mu\nu}F^{\mu\nu}+\frac{1}{2}\epsilon^{\kappa\lambda\mu\nu}A_\lambda(\hat{k}_{AF})_\kappa F_{\mu\nu}\, . \label{general cpt-odd lagrangian}
\end{equation}

The corresponding action can be written as
\begin{equation}
\mathcal{S}={}-\frac{1}{4}\int d^4x\left(F_{\mu\nu}F^{\mu\nu}-2\epsilon^{\kappa\lambda\mu\nu}A_\lambda(\hat{k}_{AF})_\kappa F_{\mu\nu}+2(\partial_\mu A^\mu)^2\right)\label{general action}
\end{equation}
where $\zeta=1$  Feynman 't Hooft gauge fixing term is used. After the surface terms are eliminated, the action is brought to the form $\mathcal{S}={}\frac{1}{2}\int d^4x A_\mu(\hat{G}^{-1})^{\mu\nu}A_\nu$. Then, the inverse propagator takes the form
\begin{equation}
(\hat{G}^{-1})^{\mu\nu}={}-\eta^{\mu\nu}(p_\sigma p^\sigma)+2i\epsilon^{\mu\kappa\lambda\nu}(\hat{k}_{AF})_\kappa p_\lambda \label{Momentum space inverse propagator}\, .
\end{equation}

From the action ($\ref{general action}$) with the adoption of the plane wave ansatz $A_\mu(x)=A_\mu(p)e^{-ix.p}$, equations of motion take the form $M^{\mu\nu}A_\nu=0$ for
\begin{equation}
M^{\mu\nu}={}\eta^{\mu\nu}p_\alpha p^\alpha-p^\mu p^\nu-2i\epsilon^{\mu\nu\alpha\beta}(\hat{k}_{AF})_\alpha p_\beta \label{general form of M}
\end{equation}
from \refe{23} of \refs{\cite{Kostelecky2009}}.
\subsection{The Dispersion Relation}
\label{s:DR}

The dispersion relation for the Lagrangian ($\ref{general cpt-odd lagrangian}$) can be obtained via the usual way, first handling the gauge fixing and then calculating the determinant of reduced linear equations. Alternatively, rank-nullity can be used to find the covariant form of dispersion relations without sacrificing the gauge invariance, as is done in \refs{\cite{Kostelecky2009}}. We use this alternative method, and obtain\footnote{See the details in \refa{\ref{App:HB}}.} from the general result \refe{1} of \refs{\cite{Kostelecky2009}}
\begin{equation}
0={}\left(p_\mu p^\mu\right)^2+4p_\alpha p^\alpha(\hat{k}_{AF})_\mu(\hat{k}_{AF})^\mu-4\left( p_\mu(\hat{k}_{AF})^\mu\right)^2 \label{General CPT-odd Dispersion Relation}\,.
\end{equation}

Special models such as vacuum, general vacuum-orthogonal and camouflage models can be most transparently applied if the spherical decomposition method is employed. To do that, we first set the helicity basis as the space part of the coordinate system. In this basis, \refe{\ref{General CPT-odd Dispersion Relation}} becomes
\begin{equation}
0={}\left(p_\mu p^\mu\right)^2-4\left(p(\hat{k}_{AF})_0-\omega(\hat{k}_{AF})_r\right)^2-8p_\mu p^\mu(\hat{k}_{AF})_+(\hat{k}_{AF})_-\label{General CPT-odd Spherical Dispersion Relation}\,,
\end{equation}
where $\omega$ is the usual frequency and $p$ denotes the magnitude of the space part of $p^\mu$.

Here, $(\hat{k}_{AF})_i$ can be expanded over spin-weighted spherical harmonics. The prescription for such an expansion is given in \refe{47-51} in \refs{\cite{Kostelecky2009}}, for which \refe{\ref{General CPT-odd Spherical Dispersion Relation}} becomes
\begin{widetext}
\begin{equation}
\begin{aligned}
0 ={}&\left(p_\mu p^\mu\right)^2-4\left(\sum\limits_{dnjm}\omega^{d-3-n}p^n\prescript{}{0}{Y}_{jm}(\mathbf{\hat{p}})\left(\frac{dp}{n+3}(k_{AF}^{(d)})^{(0B)}_{njm}+\frac{\omega}{n+2}(k_{AF}^{(d)})^{(1B)}_{njm}\right)\right)^2-8p_\mu p^\mu\\{}&\times\sum\limits_{d_1d_2n_1n_2j_1j_2m_1m_2}\omega^{d_1+d_2-6-n_1-n_2}p^{n_1+n_2}\prescript{}{+1}{Y}_{j_1m_1}(\mathbf{\hat{p}})\prescript{}{-1}{Y}_{j_2m_2}(\mathbf{\hat{p}})\frac{1}{\sqrt{4j_1j_2(j_1+1)(j_2+1)}}\\
{}&\times\left((k_{AF}^{(d_1)})^{(1B)}_{n_1j_1m_1}+i(\vok_{AF}^{(d_1)})^{(1E)}_{n_1j_1m_1}\right)\left(-(k_{AF}^{(d_2)})^{(1B)}_{n_2j_2m_2}+i(\vok_{AF}^{(d_2)})^{(1E)}_{n_2j_2m_2}\right)\,. \label{General CPT-odd Dispersion Relation2}
\end{aligned}
\end{equation}
\end{widetext}

This is the most general dispersion relation for CPT-odd {\it nm}SME for the photon sector. As it stands, it is quite complicated; however, we will show in the next section that the last term will drop so as to have a corresponding physical polarization vector.
\subsection{Polarization Vectors}
\label{s:PV}
In order to determine the photon field $A_\mu$, one needs to solve the equations of motion $M^{\mu\nu}A_\nu=0$. The necessary condition for non-trivial solution is $det(M)=0$, through which one finds the dispersion relations. The standard method is to apply these conditions on $M$ and find the corresponding polarization vectors. As extracting the generic explicit forms of the dispersion relation out of the implicit formula (\ref{General CPT-odd Dispersion Relation2}) is quite formidable, we will pursue an alternative way here. We will calculate the rank of $M$ using a generic frequency $\omega$, and obtain the constraints from the requirement $M$ having at most rank 2.\footnote{A rank-3 $M$ gives gauge solution only, and a rank-4 $M$ gives the trivial solution.} Then, these constraints will be applied to the dispersion relations, which we already worked out, in order to determine whether there exists a nontrivial coefficient subspace with a physical polarization vector obeying the general dispersion relation \refe{\ref{General CPT-odd Dispersion Relation2}}. 

In helicity basis, \refe{\ref{general form of M}} reduces to the form
\begin{equation*}
\begin{aligned}
M_\mu^{\;\nu}={}&\delta_\mu^{\;\nu}(\omega^2-p^2)-\eta_{	\mu 0}\delta^\nu_0\omega^2-\left(\eta_{\mu 0}\delta^\nu_r+\eta_{\mu r}\delta^\nu_0\right)p\omega-\eta_{\mu r}\delta^\nu_rp^2+2\left(\eta_{\mu 0}\delta^\nu_+-\eta_{\mu +}\delta^\nu_0\right)(\hat{k}_{AF})_-p\\{}&-2\left(\eta_{\mu -}\delta^\nu_0-\eta_{\mu 0}\delta^\nu_-\right)(\hat{k}_{AF})_+p-2\left(\eta_{\mu +}\delta^\nu_--\eta_{\mu -}\delta^\nu_+\right)\left((\hat{k}_{AF})_0p-(\hat{k}_{AF})_r \omega\right)\\{}&+2\left(\eta_{\mu +}\delta^\nu_r-\eta_{\mu r}\delta^\nu_+\right)(\hat{k}_{AF})_-\omega+2\left(\eta_{\mu r}\delta^\nu_--\eta_{\mu -}\delta^\nu_r\right)(\hat{k}_{AF})_+\omega
\,.\end{aligned}
\end{equation*}

Then, by the matrix representation convention $M_\rho^{\;\;\nu}$ in $(0,+,r,-)$ basis,
\begin{widetext}
\begin{equation}
M\doteq
\begin{pmatrix}
-p^2 & -2(\hat{k}_{AF})_-p & -p\omega & 2(\hat{k}_{AF})_+p\\
2(\hat{k}_{AF})_+p & \omega^2-p^2+2\left((\hat{k}_{AF})_0p+(\hat{k}_{AF})_r\omega\right) & 2(\hat{k}_{AF})_+\omega & 0\\
p\omega & 2(\hat{k}_{AF})_-\omega & \omega^2 & -2(\hat{k}_{AF})_+\omega \\
-2(\hat{k}_{AF})_-p & 0 & -2(\hat{k}_{AF})_-\omega & \omega^2-p^2-2\left((\hat{k}_{AF})_0p+(\hat{k}_{AF})_r\omega\right)
\end{pmatrix}\label{conventional rooted M in helicity basis}\,.
\end{equation}
\end{widetext}

This alternative method can be tested in no LV limit. For this case, \refe{\ref{conventional rooted M in helicity basis}} becomes
\begin{equation}
M\doteq
\begin{pmatrix}
-p^2 & 0 & -p\omega & 0\\
0 & \omega^2-p^2 & 0 & 0\\
p\omega & 0 & \omega^2 & 0 \\
0 & 0 & 0 & \omega^2-p^2
\end{pmatrix}\,. \label{No LV M without omega=p}
\end{equation}
This matrix is of rank 3 for $\omega\ne p$, which means the requirement of at most rank 2 $M$ enforces the condition $\omega=p$. Therefore, $M$ reduces to the form
\begin{equation}
M\doteq
\begin{pmatrix}
-p^2 & 0 & -p^2 & 0\\
0 & 0 & 0 & 0\\
p^2 & 0 & p^2 & 0 \\
0 & 0 & 0 & 0
\end{pmatrix}\,, \label{No LV M}
\end{equation}
for which $MA=0$ yields the following solutions
\begin{equation}
A_\mu\in\left\{\begin{pmatrix}1\\0\\-1\\0
\end{pmatrix}, \begin{pmatrix}0\\1\\0\\0
\end{pmatrix}, \begin{pmatrix}0\\0\\0\\1
\end{pmatrix}\right\}\,.\label{conventional polarization vectors}
\end{equation}

This is the expected result in Lorenz gauge. When we further apply Coulomb gauge, first component will be set to 0, which in turn kills the first solution with the radial component. Hence, there remain two transverse solutions with the same dispersion relation $\omega=p$.

The same procedure can be utilized for the general case with the Lorentz violation. From \refe{\ref{conventional rooted M in helicity basis}}, the equations of motion become
\begin{widetext}
\begin{equation}
\begin{pmatrix}
-p^2 & -2(\hat{k}_{AF})_-p & -p\omega & 2(\hat{k}_{AF})_+p\\
2(\hat{k}_{AF})_+p & \omega^2-p^2-2p(\hat{k}_{AF})_s & 2(\hat{k}_{AF})_+\omega & 0\\
p\omega & 2(\hat{k}_{AF})_-\omega & \omega^2 & -2(\hat{k}_{AF})_+\omega \\
-2(\hat{k}_{AF})_-p & 0 & -2(\hat{k}_{AF})_-\omega & \omega^2-p^2+2p(\hat{k}_{AF})_s
\end{pmatrix}\begin{pmatrix}
A_0 \\ -A_+ \\ -A_r \\ -A_-
\end{pmatrix}=\begin{pmatrix}
0\\0\\0\\0
\end{pmatrix}\,, \label{Equations of Motion for photon field}
\end{equation}
\end{widetext}
where the minus signs in $A$ are because $A$ should be in covariant form as $MA=0$ reads $M_\mu^{\;\nu}A_\nu=0$ due to the matrix convention used. Also we define
\begin{equation}
(\hat{k}_{AF})_s:={}(\hat{k}_{AF})_0+\frac{\omega}{p}(\hat{k}_{AF})_r \label{definition of k_s}
\end{equation}
for brevity.

As stated earlier, the rank of $M$ should be at most 2 for physical solutions to emerge. One can show that there are various possible combinations of LVT which assures that this rank condition is satisfied. Analyzing all possibilities can be quite lengthy, and are assumed to give little insight; therefore, we choose one of the possibilities and analyze it further.

The LV coefficients are restricted such that
\begin{equation}
\begin{aligned}
(\hat{k}_{AF})_+={}& 0\,,\\
(\hat{k}_{AF})_-={}& 0\,.
\end{aligned}
\end{equation}
which gives a resultant $M$
\begin{widetext}
\begin{equation}
\begin{pmatrix}
-p^2 & 0 & -p\omega & 0\\
0 & \omega^2-p^2-2p(\hat{k}_{AF})_s & 0 & 0\\
p\omega & 0 & \omega^2 & 0 \\
0 & 0 & 0 & \omega^2-p^2+2p(\hat{k}_{AF})_s
\end{pmatrix}
\end{equation}
\end{widetext}
which is clearly at most rank 2 when the dispersion relation is imposed.

One of the advantages of this coefficient form is that the dispersion relation is not as formidable as it would be in the case of nonzero $(\hat{k}_{AF})_\pm$. The second advantage is that these restrictions kill the last term on the \refe{\ref{General CPT-odd Dispersion Relation2}}.

The coefficient subspace under this restriction further splits into two parts regarding $(\hat{k}_{AF})_s=0$ or not. It turns out that the polarization vectors and the dispersion relation which they obey become conventional if $(\hat{k}_{AF})_s=0$, whereas there arises two transverse polarization vectors with two different dispersion relations if  $(\hat{k}_{AF})_s\ne0$.

Concisely, the generic coefficient space for nonrenormalizable CPT-odd extension of photon sector can be divided into three parts as in \reft{\ref{Table-cscops}}. Any analysis with a definite set of nonzero $\hat{k}_{AF}$ coefficients, depending on the relation between the components of the coefficients, necessarily falls into one of these three groups: One with conventional solutions, one with birefringent solutions and one which is not analyzed in this paper. For later convenience, we denote these coefficient subspaces with $\hat{k}_{AF}^{(cn)}$, $\hat{k}_{AF}^{(bf)}$, and $\hat{k}_{AF}^{(na)}$ where $cn$, $bf$,  and $na$ refer to nature of resultant polarization vectors: conventional, birefringent, and not analyzed, respectively.

It should also be noted that the conditions that define the coefficient subspaces and any conclusions drawn from them are not valid at all observer frames. This is because the components of LVT are mixed up under observer transformations, i.e., $\hat{k}_{AF}$ transforms as a four vector under observer Lorentz transformation. Therefore, one should not view \reft{\ref{Table-cscops}} as an invariantly valid classification of all possible $\hat{k}_{AF}$, but instead as a tool to deduce the properties of the modified photon after a specific observer frame and nonzero components of LVT are selected.

There may be a subtle point that the conditions of having zero components for $\hat{k}_{AF}$ would be ambiguous. Does requiring zero components for $\hat{k}_{AF}$ in one helicity basis lead to a null $\hat{k}_{AF}$ in another basis? Since basis vectors are dependent on the photon direction which is arbitrary in a priori chosen observer frame of any analysis, one can change the basis vectors by choosing different photon directions, which is equivalent to the so-called particle Lorentz transformations. However, unlike under observer Lorentz transformations, $\hat{k}_{AF}$ do not transform as four vector under particle Lorentz transformations; instead, it transforms as set of scalars. Therefore, assumption of some zero $\hat{k}_{AF}$ components regardless of photon momentum direction is unambiguous and valid.

\begin{table*}
\caption{\label{Table-cscops}Coefficient subspace of CPT-odd modified photon sector with physical solutions}
\begin{ruledtabular}
\begin{tabularx}{\textwidth}{@{}c c c c@{}}
Coefficient Subspace & Conditions & Dispersion Relation & Polarization Vectors \\ \hline\hline
$\hat{k}_{AF}^{(bf)}$
&$\begin{array}{r l}
(\hat{k}_{AF})_+={}& 0\\(\hat{k}_{AF})_-={}& 0\\(\hat{k}_{AF})_0-(\hat{k}_{AF})_r \ne{}& 0
\end{array}$&$\begin{array}{l}
\omega^2-p^2-2p(\hat{k}_{AF})_s={}0\\\omega^2-p^2+2p(\hat{k}_{AF})_s={}0
\end{array}$& $\begin{array}{c}
\left\{\begin{pmatrix}
\omega\\0\\p\\0
\end{pmatrix} , \begin{pmatrix}
0\\0\\0\\1
\end{pmatrix}\right\}\\\left\{\begin{pmatrix}
\omega\\0\\p\\0
\end{pmatrix} , \begin{pmatrix}
0\\1\\0\\0
\end{pmatrix}\right\}
\end{array}$\\\hline
$\hat{k}_{AF}^{(cn)}$
&$\begin{array}{r l}
(\hat{k}_{AF})_-={}& 0\\(\hat{k}_{AF})_+={}& 0\\(\hat{k}_{AF})_0-(\hat{k}_{AF})_r={}& 0
\end{array}$&$\omega=p$&$\left\{\begin{pmatrix}
1\\0\\1\\0
\end{pmatrix}, \begin{pmatrix}
0\\1\\0\\0
\end{pmatrix}, \begin{pmatrix}
0\\0\\0\\1
\end{pmatrix}\right\}$\\\hline
$\hat{k}_{AF}^{(na)}$
&$\{(\hat{k}_{AF})_+\ne0\} \lor \{(\hat{k}_{AF})_-\ne0\}$ & \refe{\ref{General CPT-odd Dispersion Relation2}} & \text{Not Analyzed}
\end{tabularx}
\end{ruledtabular}
\end{table*}

\subsection{The Propagator}
\label{s:P}
The task of analytical inversion of the inverse propagator ($\ref{Momentum space inverse propagator}$) while remaining covariant is formidable, if possible. One can assume a leading order ansatz or a perturbation expansion so as to preserve the covariant form; however, since we are already interested in a specific form of LV in the Helicity basis, it would be more appropriate to derive the propagator in the Helicity basis itself.This method is also advantageous as one can always go to the matrix representation once an explicit basis is chosen, and the non-singular inverse propagator matrix can always be analytically inverted; hence, we obtain the exact form of the propagator without an approximation. 

The inverse propagator (\ref{Momentum space inverse propagator}) can further be decomposed\footnote{See the details in \refa{\ref{App:HB}}.} and be brought to the form
\begin{equation}
\begin{aligned}
(\hat{G}^{-1})^{\mu\nu}={}&-\eta^{\mu\nu}(p_\sigma p^\sigma)+2\delta^\mu_0\delta^\nu_+p(\hat{k}_{AF})_--2\delta^\mu_+\delta^\nu_0p(\hat{k}_{AF})_--2\delta^\mu_0\delta^\nu_-p(\hat{k}_{AF})_++2\delta^\mu_-\delta^\nu_0p(\hat{k}_{AF})_+\\{}
&-2\delta^\mu_-\delta^\nu_+(\omega(\hat{k}_{AF})_r+p(\hat{k}_{AF})_0)+2\delta^\mu_+\delta^\nu_-(\omega(\hat{k}_{AF})_r+p(\hat{k}_{AF})_0)-2\delta^\mu_r\delta^\nu_-\omega(\hat{k}_{AF})_+\\{}
&+2\delta^\mu_-\delta^\nu_r\omega(\hat{k}_{AF})_+-2\delta^\mu_+\delta^\nu_r\omega(\hat{k}_{AF})_-+2\delta^\mu_r\delta^\nu_+\omega(\hat{k}_{AF})_-
\end{aligned}\label{Equation: Helicity-basis inverse propagator}
\end{equation}
which reduces to
\begin{equation}
\begin{aligned}
(\hat{G}^{-1})^{\mu\nu}={}-\eta^{\mu\nu}(p_\sigma p^\sigma)-2\delta^\mu_-\delta^\nu_+(\omega(\hat{k}_{AF})_r+p(\hat{k}_{AF})_0)+2\delta^\mu_+\delta^\nu_-(\omega(\hat{k}_{AF})_r+p(\hat{k}_{AF})_0)
\end{aligned}
\end{equation}
in the interested coefficient regime, and hence it can be represented in the matrix representation as
\begin{widetext}
\begin{equation}
\begin{aligned}
(\hat{G}^{-1})\doteq{}
\begin{pmatrix}
-(p_\sigma p^\sigma)& 0& 0& 0\\
0&-(p_\sigma p^\sigma)+2p(\hat{k}_{AF})_s& 0& 0\\
0& 0&-(p_\sigma p^\sigma)& 0\\
0& 0& 0&-(p_\sigma p^\sigma)-2p(\hat{k}_{AF})_s
\end{pmatrix}\,.\label{inverse propagator in helicity}
\end{aligned}
\end{equation}
\end{widetext}

Therefore,
\begin{equation}
\begin{aligned}
\hat{G}\doteq{}\text{diagonal}\left(-\frac{1}{(p_\sigma p^\sigma)},-\frac{1}{(p_\sigma p^\sigma)-2p(\hat{k}_{AF})_s},-\frac{1}{(p_\sigma p^\sigma)}, -\frac{1}{(p_\sigma p^\sigma)+2p(\hat{k}_{AF})_s}\right)\,.
\end{aligned}
\end{equation}

\section{The CPT-odd Vacuum-Orthogonal Model}
\label{sec:VOMFCOP}
When the {\it nm}SME photon sector and the decompose of related LVT over spin-weighted spherical harmonics were introduced in the paper \refs{\cite{Kostelecky2009}}, the possibility of specialized models and their constructions were also presented and discussed. As stressed out in the \refp{\ref{intro}}, the main advantageous of helicity basis is its relevance to direct observation, which makes a decomposition in this basis decouple LVT according to their observable effects.

A directly relevant observable effect of LVT is that on the vacuum propagation. If one restricts the attention to only those LVT which generate leading order dispersive or birefringence effects on vacuum, then the associated model is named \emph{vacuum model}. On the contrary, if one restricts the attention to the coefficient subspace, which is compliment of that of vacuum model, then the associated model is called \emph{vacuum orthogonal model}.

In \refp{\ref{sec:COPS}}, we analyzed the coefficient space with the simple coefficient set of \refe{\ref{Eq:decompose of k}}, which does not differ vacuum properties. However, one needs to consider only vacuum orthogonal LVT for a vacuum orthogonal model; hence, the conversion from this simple set to those in \reft{\ref{Table-sdc}}, followed by the impose of $k_{(V)jm}^{(d)}=0$ is required. Fortunately, there is a simple prescription for this conversion, given by \refe{97-98} of \refs{\cite{Kostelecky2009}}.

\subsection{Dispersion Relation and Polarization Vectors}
\label{s:VODR}
Being a special case of CPT-odd sector, the vacuum orthogonal model obeys the general CPT-odd dispersion relation (\ref{General CPT-odd Dispersion Relation2}). The mere modification is the application of prescription mentioned above, for which the dispersion relation reduces to the form
\begin{widetext}
\begin{equation}
\begin{aligned}
0={}&\left(p_\mu p^\mu\right)^2-4\Bigg\{\sum\limits_{dnjm}\omega^{d-3-n}p^n\prescript{}{0}{Y}_{jm}(\mathbf{\hat{p}})\Bigg[(\vok_{AF}^{(d)})^{(0B)}_{njm}\frac{p}{\omega^2}\left(\frac{(d-2-n)\omega^2}{d-2-n+j}-\frac{(d-4-n)p^2}{d-4-n+j}\right)\nonumber\\{}&-\frac{dp^2}{(n+4)(n+2)\omega}(\vok_{AF}^{(d)})^{(1B)}_{njm}+(\vok_{AF}^{(d)})^{(0B)}_{njm}\frac{j}{p}\left(\frac{\omega^2}{d-2-n+j}-\frac{p^2}{d-4-n+j}\right)\nonumber\\{}&+\frac{d\omega}{(n+2)(n+4)}(\vok_{AF}^{(d)})^{(1B)}_{njm}\Bigg]\Bigg\}^2+8p_\mu p^\mu\sum\limits_{d_1d_2n_1n_2j_1j_2m_1m_2}\omega^{d_1+d_2-6-n_1-n_2}p^{n_1+n_2}\prescript{}{+1}{Y}_{j_1m_1}(\mathbf{\hat{p}})\nonumber\\{}
&\times\prescript{}{-1}{Y}_{j_2m_2}(\mathbf{\hat{p}})\frac{1}{\sqrt{4j_1j_2(j_1+1)(j_2+1)}}\Bigg\{\Bigg[\left(\frac{\omega j_1(n_1+1)}{p(d_1-2-n_1+j_1)}-\frac{pj_1(n_1+3)}{\omega(d_1-4-n_1+j_1)}\right)(\vok_{AF}^{(d_1)})_{n_1j_1m_1}^{(0B)}\nonumber\\{}&+\frac{d_1}{n_1+4}(\vok_{AF}^{(d_1)})^{(1B)}_{n_1j_1m_1}\Bigg]\Bigg[\left(\frac{\omega j_2(n_2+1)}{p(d_2-2-n_2+j_2)}-\frac{pj_2(n_2+3)}{\omega(d_2-4-n_2+j_2)}\right)(\vok_{AF}^{(d_2)})_{n_2j_2m_2}^{(0B)}\nonumber\\{}&+\frac{d_2}{n_2+4}(\vok_{AF}^{(d_2)})^{(1B)}_{n_2j_2m_2}\Bigg]+(\vok_{AF}^{(d_1)})^{(1E)}_{n_1j_1m_1}(\vok_{AF}^{(d_2)})^{(1E)}_{n_2j_2m_2}\Bigg\}\,.\label{General Vacuum-orthogonal dispersion relation}
\end{aligned}
\end{equation}
\end{widetext}

As it stands, this dispersion relation does not give much insight. However, it can be cast into the form\footnote{See the details in \refa{\ref{App:Covodr}}.}
\begin{equation}
0={}(p_\mu p^\mu)\times\Big((p_\mu p^\mu)\mathcal{P}(\omega,p)+\mathcal{Q}(\omega,p)\Big)\,,\label{General Vacuum-orthogonal dispersion relation2}
\end{equation}
where $\mathcal{P}$ and $\mathcal{Q}$ are defined as
\begin{widetext}
\begin{subequations}
\begin{align}
\mathcal{P}(\omega,p):={}& 1-4\Bigg\{\sum\limits_{dnjm}\omega^{d-4-n}p^n\prescript{}{0}{Y}_{jm}(\mathbf{\hat{p}})\left((\vok_{AF}^{(d)})^{(1B)}_{njm}\frac{d}{(n+2)(n+4)}\right)\nonumber\\{}&+\omega^{d-5-n}p^{n-1}\prescript{}{0}{Y}_{jm}(\mathbf{\hat{p}})(\vok_{AF}^{(d)})^{(0B)}_{njm}\left(\omega^2\frac{j}{d-2-n+j}+p^2\frac{d-4-n}{d-4-n+j}\right)\Bigg\}^2\,,\\
\mathcal{Q}(\omega,p):={}& 8\sum\limits_{d_1d_2n_1n_2j_1j_2m_1m_2}\omega^{d_1+d_2-6-n_1-n_2}p^{n_1+n_2}\prescript{}{+1}{Y}_{j_1m_1}(\mathbf{\hat{p}})\prescript{}{-1}{Y}_{j_2m_2}(\mathbf{\hat{p}})\frac{1}{\sqrt{4j_1j_2(j_1+1)(j_2+1)}}\nonumber\\
{}&\times\Bigg\{\Bigg(\left(\frac{\omega j_1(n_1+1)}{p(d_1-2-n_1+j_1)}-\frac{pj_1(n_1+3)}{\omega(d_1-4-n_1+j_1)}\right)(\vok_{AF}^{(d_1)})_{n_1j_1m_1}^{(0B)}+\frac{d_1}{n_1+4}(\vok_{AF}^{(d_1)})^{(1B)}_{n_1j_1m_1}\Bigg)\nonumber\\{}&\times\Bigg(\left(\frac{\omega j_2(n_2+1)}{p(d_2-2-n_2+j_2)}-\frac{pj_2(n_2+3)}{\omega(d_2-4-n_2+j_2)}\right)(\vok_{AF}^{(d_2)})_{n_2j_2m_2}^{(0B)}+\frac{d_2}{n_2+4}(\vok_{AF}^{(d_2)})^{(1B)}_{n_2j_2m_2}\Bigg)\nonumber\\{}&+(\vok_{AF}^{(d_1)})^{(1E)}_{n_1j_1m_1}(\vok_{AF}^{(d_2)})^{(1E)}_{n_2j_2m_2}\Bigg\}\,.
\end{align}\label{P and Q of dispersion relation}
\end{subequations}
\end{widetext}

In our interested coefficient range, subject to $(\hat{k}_{AF})_\pm=0$ restriction, it turns out that $\mathcal{Q}(\omega,p)=0$ for the vacuum orthogonal subspace. Then \refe{\ref{General Vacuum-orthogonal dispersion relation2}} becomes
\begin{equation}
0={}\left(p_\mu p^\mu\right)^2\left(1+\mathcal{R}(\omega,p)\right)\left(1-\mathcal{R}(\omega,p)\right)\,,\label{General Vacuum-orthogonal dispersion relation3}
\end{equation}
where $\mathcal{R}$ is defined as
\begin{equation}
\begin{aligned}
\mathcal{R}(\omega,p):={}& 2\sum\limits_{dnjm}\Bigg\{\omega^{d-4-n}p^n\prescript{}{0}{Y}_{jm}(\mathbf{\hat{p}})\left((\vok_{AF}^{(d)})^{(1B)}_{njm}\frac{d}{(n+2)(n+4)}\right)\\{}&+\omega^{d-5-n}p^{n-1}\prescript{}{0}{Y}_{jm}(\mathbf{\hat{p}})(\vok_{AF}^{(d)})^{(0B)}_{njm}\left(\omega^2\frac{j}{d-2-n+j}+p^2\frac{d-4-n}{d-4-n+j}\right)\Bigg\}\,. \label{definition of R}
\end{aligned}
\end{equation}

The dispersion relation in \refe{\ref{General Vacuum-orthogonal dispersion relation3}} has three roots: $\omega=p$ and $\mathcal{R}(\omega,p)\pm1=0$. As can be clearly deduced from \reft{\ref{Table-cscops}}, the first root is the conventional dispersion relation that the selection of nonzero $\hat{k}_{AF}$ components lying on the coefficient subspace $\hat{k}_{AF}^{(cn)}$ gives rise, and the other two are the dispersion relations that the birefringent solutions, which a selection of $\hat{k}_{AF}\in\hat{k}_{AF}^{(bf)}$ gives rise, obey. However, to prevent any ambiguity, the dispersion relations for the specific cases will be explicitly calculated below.

In the vacuum orthogonal coefficient subspace, the term \mbox{$p(\hat{k}_{AF})_s=p(\hat{k}_{AF})_0+\omega(\hat{k}_{AF})_r$} for $\hat{k}_{AF}^{(b)}$ becomes
\begin{widetext}
\begin{equation*}
\begin{aligned}
p(\hat{k}_{AF})_0+\omega(\hat{k}_{AF})_r={}& (p_\mu p^\mu)\Bigg\{\sum\limits_{dnjm}\omega^{d-4-n}p^n\prescript{}{0}{Y}_{jm}(\mathbf{\hat{p}})\Bigg((\vok_{AF}^{(d)})^{(1B)}_{njm}\frac{d}{(n+2)(n+4)}\Bigg)\nonumber\\{}&+\omega^{d-5-n}p^{n-1}\prescript{}{0}{Y}_{jm}(\mathbf{\hat{p}})(\vok_{AF}^{(d)})^{(0B)}_{njm}\left(\omega^2\frac{j}{d-2-n+j}+p^2\frac{d-4-n}{d-4-n+j}\right)\Bigg\}\,;
\end{aligned}
\end{equation*}
\end{widetext}
hence,
\begin{equation*}
p(\hat{k}_{AF})_s={}\frac{1}{2}(p_\mu p^\mu)\mathcal{R}(\omega,p)\,.
\end{equation*}

Once this is inserted into the the dispersion relations in \reft{\ref{Table-cscops}}, they become
\begin{equation*}
\omega^2-p^2\pm 2p(\hat{k}_{AF})_s=0\quad \longrightarrow\quad (p_\mu p^\mu)\left(1\pm\mathcal{R}(\omega,p)\right)=0\,.
\end{equation*}

For $\omega=p$ root, $(\hat{k}_{AF})_s$ is forced to be zero which dictates $(\hat{k}_{AF})_s=0$. However, this contradicts with the requirement of $\hat{k}_{AF}^{(bf)}$ in which these dispersion relations are valid. Therefore, the only roots of the dispersion relation for $\hat{k}_{AF}^{(bf)}$ are \mbox{$\mathcal{R}(\omega,p)\pm1=0$}, which is exactly our earlier deduction.

At the first glance, there seems a contradiction in the results. The vacuum orthogonal coefficient subspace should not have produced birefringent results; after all, the name vacuum orthogonal asserts no leading order vacuum birefringence. The results are consistent though, as the birefringent dispersion relations $\mathcal{R}(\omega,p)\pm1=0$ are not actually so-called \emph{perturbative solutions}, which smoothly reduces to the conventional dispersion relation as Lorentz violation is switched off. They are so-called \emph{spurious solutions} \cite{Shreck2013}, which blow up as LVT go to zero. According to \refs{\cite{Kostelecky2009}}, these solutions are Planck scale effects and should be neglected. We will explicitly show that these solutions blow up in \refp{\ref{sec:SMA}}.

The resultant situation is actually worth repetition. In \refp{\ref{s:PV}}, we show that the generic coefficient space of all possible analysis of nonrenormalizable CPT-odd extension of the photon sector can be divided into three coefficient subspaces: $\hat{k}_{AF}^{(bf)}$, $\hat{k}_{AF}^{(cn)}$, and $\hat{k}_{AF}^{(na)}$. Therefore, once an observer frame is chosen and a particular nonzero coefficient set for that frame is taken among the components of $\hat{k}_{AF}$, then the resultant properties can be directly read off from \reft{\ref{Table-cscops}}, which allows the conventional solutions to emerge despite the presence of some specific LVT. For the general vacuum orthogonal model, $\hat{k}_{AF}^{(bf)}$  becomes irrelevant as it produces merely spurious solutions; hence, the conventional solutions are the only physical solutions for the vacuum orthogonal model with the chosen LVT. Since the model has no birefringent solutions at any order\footnote{As construction, the vacuum orthogonal model should not have leading order birefringence, but that by no means prevents it to have higher order birefringence effects.}, we say that, for this LVT, \emph{the vacuum orthogonal model is vacuum orthogonal at all orders}, and \emph{all polarization vectors and their dispersion relations remain conventional in vacuum orthogonal model}.

\subsection{The Coefficient Subspace $\hat{k}_{AF}^{(cn)}$ in Vacuum Orthogonal Model}
\label{s:VOPV}
The vacuum orthogonal model at hand has physical and relevant solutions only in the coefficient subspace $\hat{k}_{AF}^{(cn)}$ as shown in \refp{\ref{s:VODR}}. Because $\hat{k}_{AF}^{(cn)}$ is defined as the subspace for which $(\hat{k}_{AF})_\pm=0$ and $(\hat{k}_{AF})_s=0$ hold, the relevant coefficient subspace of vacuum orthogonal model is the vacuum orthogonal version of these constraints. One can show that these equations translate into the following conditions in the vacuum orthogonal coefficient subspace.
\begin{widetext}
\textbf{$\bm{(\hat{k}_{AF})_0+(\hat{k}_{AF})_r}=0$ Condition:}
\begin{equation}
\begin{aligned}
0={}&\sum\limits_{n}(\vok_{AF}^{(d)})^{(0B)}_{njm}\left(-\frac{4}{d}+\frac{4j(d+1+j)}{d(d-2-n+j)(d-4-n+j)}\right)\\
{}&-\sum\limits_{n}(\vok_{AF}^{(d)})^{(1B)}_{njm}\left(\frac{1}{n+2}-\frac{d}{(n+4)(n+2)}-\frac{(d-3-n)(n+4)}{d(d-3-n+j)(n+2)}\right)\label{first condition for vacuum polarization}
\end{aligned}
\end{equation}

\textbf{$\bm{(\hat{k}_{AF})_\pm=0}$ Condition:}
\begin{equation}
\begin{aligned}
\sum\limits_{n}\left(-\frac{2j(d-1+j)}{(d-2-n+j)(d-4-n+j)}(\vok_{AF}^{(d)})^{(0B)}_{njm}+\frac{d}{n+4}(\vok_{AF}^{(d)})^{(1B)}_{njm}\right)={}& 0\,, \\
\sum\limits_{n}(\vok_{AF}^{(d)})^{(1E)}_{njm}={}& 0
\end{aligned}\label{second condition for vacuum polarization}
\end{equation}
\end{widetext}

\begin{turnpage}
\begin{table*}
\caption{\label{Table-vos} Vacuum Orthogonal Solutions.}
\begin{ruledtabular}
\begin{tabular}{c c c c}
\textbf{Coefficient Subspace} & Conditions & \textbf{Dispersion Relation} & \textbf{Polarization Vectors} $A^\mu$\\\hline
$\hat{k}_{AF}^{(cn)}$ & $\begin{aligned}
0={}&\sum\limits_{n}(\vok_{AF}^{(d)})^{(0B)}_{njm}\left(-\frac{4}{d}+\frac{4j(d+1+j)}{d(d-2-n+j)(d-4-n+j)}\right)\\
{}&+\sum\limits_{n}(\vok_{AF}^{(d)})^{(1B)}_{njm}\bigg(\frac{1}{n+2}-\frac{d}{(n+4)(n+2)}\\
{}&-\frac{(d-3-n)(n+4)}{d(d-3-n+j)(n+2)}\bigg)\\
0={}&\sum\limits_{n}\bigg(-\frac{2j(d-1+j)}{(d-2-n+j)(d-4-n+j)}(\vok_{AF}^{(d)})^{(0B)}_{njm}\\
&{}+\frac{d}{n+4}(\vok_{AF}^{(d)})^{(1B)}_{njm}\bigg)\\
0={}&\sum\limits_{n}(\vok_{AF}^{(d)})^{(1E)}_{njm}
\end{aligned}$
& $\omega=p$ & $\left\{\begin{pmatrix}
1\\0\\1\\0
\end{pmatrix},\begin{pmatrix}
0\\1\\0\\0
\end{pmatrix},\begin{pmatrix}
0\\0\\0\\1
\end{pmatrix}\right\}$\\\hline
$\hat{k}_{AF}^{(bf)}$ & Given by compliment LVT combinations under $(\hat{k}_{AF})_\pm=0$
 & Spurious & Physically Irrelevant
\end{tabular}
\end{ruledtabular}
\end{table*}
\end{turnpage}

\subsection{The Propagator}
In the vacuum orthogonal model, the general {\it nm}SME CPT-odd modified photon propagator can be further simplified as follows
\begin{equation*}
\hat{G}_\mu^{\;\nu}={}-\frac{\delta_\mu^{\;\nu}}{(p_\sigma p^\sigma)}+\delta_\mu^{\;+}\delta_+^{\;\nu}\left(\frac{1}{(p_\sigma p^\sigma)}-\frac{1}{(p_\sigma p^\sigma)\left(1+\mathcal{R}(\omega,p)\right)}\right)+\delta_\mu^{\;-}\delta_-^{\;\nu}\left(\frac{1}{(p_\sigma p^\sigma)}-\frac{1}{(p_\sigma p^\sigma)\left(1-\mathcal{R}(\omega,p)\right)}\right) 
\end{equation*}
from the equivalence $ 2p(\hat{k}_{AF})_s=(p_\mu p^\mu)\mathcal{R}(\omega,p)$ in the vacuum orthogonal coefficient subspace, which is showed in \refp{\ref{s:VOPV}}. For notational convenience, we can combine these terms and rewrite as
\begin{equation}
\hat{G}_\mu^{\;\nu}={}-\frac{\delta_\mu^{\;\nu}}{(p_\sigma p^\sigma)}+\frac{1}{(p_\sigma p^\sigma)}\left(\delta_\mu^{\;+}\delta_+^{\;\nu}\frac{\mathcal{R}(\omega,p)}{1+\mathcal{R}(\omega,p)}-\delta_\mu^{\;-}\delta_-^{\;\nu}\frac{\mathcal{R}(\omega,p)}{1-\mathcal{R}(\omega,p)}\right)\,.\label{eq: General vacuum orthogonal propagator}
\end{equation}

This is the general form of the propagator for the vacuum orthogonal model. However, it contains redundant generality as the only physical solutions emerge from $\hat{k}_{AF}^{(cn)}$. We can restrict $\mathcal{R}(\omega,p)$ to this case by taking $(\hat{k}_{AF})_r$ to $(\hat{k}_{AF})_0$:
\begin{equation}
\begin{aligned}
\mathcal{R}(\omega,p)&=\frac{2p(\hat{k}_{AF})_s}{(p_\mu p^\mu)}=2\frac{p(\hat{k}_{AF})_0-\omega(\hat{k}_{AF})_r}{\omega^2-p^2}\,,\\
\lim\limits_{(\hat{k}_{AF})_r\rightarrow(\hat{k}_{AF})_0}\mathcal{R}(\omega,p)&=-\frac{2(\hat{k}_{AF})_0}{\omega+p}\,.
\end{aligned}\nonumber
\end{equation}
Then, \refe{\ref{eq: General vacuum orthogonal propagator}} becomes
\begin{equation}
\hat{G}_\mu^{\;\nu}={}-\frac{\delta_\mu^{\;\nu}}{(p_\sigma p^\sigma)}-\frac{1}{(p_\sigma p^\sigma)}\left(\delta_\mu^{\;+}\delta_+^{\;\nu}\frac{2(\hat{k}_{AF})_0}{\omega+p-2(\hat{k}_{AF})_0}-\delta_\mu^{\;-}\delta_-^{\;\nu}\frac{2(\hat{k}_{AF})_0}{\omega+p+2(\hat{k}_{AF})_0}\right)\,.\label{eq: Physical vacuum orthogonal propagator}
\end{equation}
\section{Special Model Analysis}
\label{sec:SMA}
\subsection{Vacuum Orthogonal and Isotropic Models at All Orders}
\label{ss:VOI}
The examination of a Lorentz violating model with the full LVT set is theoretically quite cumbersome and experimentally not practical. This makes working with special models inevitable, among whom vacuum and vacuum orthogonal models are introduced in \refp{\ref{sec:VOMFCOP}}. 

Another special model that can be considered is so called \emph{isotropic model}, which is also referred as ``fried-chicken" model. In such a model, all LVT except the isotropic ones are accepted to vanish in the preferred observer frame. Here, the selection of the reference frame is crucial as the vanishing terms are not necessarily zero in other reference frames. From the theoretical point of view one natural choice is the frame of \emph{Cosmic Microwave Background} (CMB) as indicated in \refs{\cite{Kostelecky2009}}. Another possible choice is the canonical Sun-centered frame, which exploits the isotropic features of the theory better for an experimental point of view.

Although isotropic models are special models in their own rights, what is examined here is only a hybrid model of vacuum orthogonal and isotropic models, where isotropic model is considered as a limiting case of general vacuum orthogonal model. That is a useful limiting case because isotropic models are somewhat popular, and moreover the result of \refp{\ref{s:VODR}} that there is no nonconventional root of dispersion relation in vacuum orthogonal models in the interested LVT regime can be better seen in this limit.

With the spherical decomposition, the condition of being isotropic translates into $j=0$, leading all LVT except $(\vok_{AF}^{(d)})^{(0B)}_{n00}$ to vanish in the CPT-odd {\it nm}SME for the photon sector as can be seen in \reft{\ref{Table-irfvoc}}. If we apply this to the general CPT-odd vacuum orthogonal dispersion relation \refe{\ref{General Vacuum-orthogonal dispersion relation2}}, the dispersion relation takes the form
\begin{equation}
0={}\left(1-\frac{p^2}{\pi}\left(\sum\limits_{d=\text{odd}>3}\sum\limits_{n=\text{even}\ge 0}^{d-5}\omega^{d-5-n}p^n\xi_{dn}\right)^2\right)(p_\mu p^\mu)^2\,, \label{vacuum orthogonal & isotropic dispersion relation}
\end{equation}
where
\begin{equation*}
\xi_{dn}:={}(\vok_{AF}^{(d)})^{(0B)}_{n00}
\end{equation*}
is defined for brevity. 

In the leading order, the dispersion relation reduces to the form
\begin{equation}
0={}\left(1-\frac{(\xi_{50} p)^2}{\pi}\right)(p_\mu p^\mu)^2 \label{vod in 5}
\end{equation}
where the Lorentz violation is purely multiplicative, and there are only two conventional roots. The multiplicative term possesses no roots for $\omega$ and is practically irrelevant since the framework is EFT which is expected to hold only for $\lvert\xi_{dn}p\rvert\ll1$.

In the next-to-leading order, the dispersion relation becomes
\begin{equation*}
0=\left(p_\mu p^\mu\right)^2\left(1+\frac{p}{\sqrt{\pi}}\left(\omega^2\xi_{70}+p^2\xi_{72}\right)\right)\left(1-\frac{p}{\sqrt{\pi}}\left(\omega^2\xi_{70}+p^2\xi_{72}\right)\right)
\end{equation*}
which can be directly compared with the general case \refe{\ref{General Vacuum-orthogonal dispersion relation3}}. The deduction there that nonconventional roots are spurious is explicitly proven here in this limit, as these roots are
\begin{equation*}
\omega^2=\pm\frac{\sqrt{\pi}}{\xi_{70}p}-\frac{\xi_{72}}{\xi_{70}}p^2\,,
\end{equation*}
which blow up as Lorentz violation is turned off. As higher orders are considered, there will be extra perturbative terms added with higher orders of $p$; however, the first term, which causes the spurious nature, will remain. 

In isotropic model analysis, the so-called \emph{ring coefficients} are preferred over the general coefficients employed above. We did the calculations in the usual coefficients as they are more transparent; however, a similar treatment can be done via the ring coefficients. In these experimentally more convenient coefficients, \refe{\ref{vacuum orthogonal & isotropic dispersion relation}} reduces to the form\footnote{See the details in \refa{\ref{App:COVOIDRIRC}}.}
\begin{equation}
0={}\Bigg\{1-\frac{p^2}{\pi}\bigg(\sum\limits_{d=\text{odd}}\sum\limits_{n=\text{even}\ge 0}^{d-5}\sum\limits_{i=\text{even}\ge 0}^{(d-5-n)}\frac{d}{n+3}\omega^{d-5-n-i}p^{i}\left((\ring{k}_{AF}^{(d)})_np^{n}\right)\bigg)^2\Bigg\}(p_\mu p^\mu)^2\,. \label{disperion relation in ring coefficients}
\end{equation}

\subsection{The Leading Order Vacuum Orthogonal Model}
\label{ss:LOVOM}
In \refp{\ref{ss:VOI}}, the isotropic limit of the general vacuum orthogonal model is considered. In this section, leading order limit $d=5$ of general vacuum orthogonal model will be examined for the similar purposes: explicit analysis of spurious roots and relevant coefficients' determination.

The dispersion relation for this leading order model is readily given by Equation (\ref{General Vacuum-orthogonal dispersion relation3}). One, then, needs to restrict \refe{\ref{definition of R}} to $d=5$ and expand it explicitly in terms of the relevant LV coefficients. However, in order to derive the more generic form of the dispersion relation which also applies to the case $\hat{k}_{AF}\in\hat{k}_{AF}^{(na)}$\footnote{We set this possibility beyond the scope of this paper in the beginning; nonetheless, it is not hard to consider this possibility for the current analysis, and it may be valuable for future research.}, the most general dispersion relation \refe{\ref{General Vacuum-orthogonal dispersion relation}} is the starting point to account for additional terms $(\hat{k}_{AF})_\pm$.

Once $d=5$ is set, \refe{\ref{General Vacuum-orthogonal dispersion relation}} becomes
\begin{widetext}
\begin{equation}
\begin{aligned}
0={}(p_\mu p^\mu)\times\Bigg\{{}&\left(p_\mu p^\mu\right)-4(p_\mu p^\mu)\Bigg[\sum_{jm}\prescript{}{0}{Y}_{jm}(\mathbf{\hat{p}})\bigg((\vok_{AF}^{(5)})_{0jm}^{(0B)}p+\frac{\omega}{3}(\vok_{AF}^{(5)})_{1jm}^{(0B)}+\frac{5\omega}{8}(\vok_{AF}^{(5)})_{0jm}^{(1B)}\\{}&+\frac{p}{3}(\vok_{AF}^{(5)})_{1jm}^{(1B)}\bigg)\Bigg]^2+8\sum\limits_{n_1n_2j_1j_2m_1m_2}\omega^{4-n_1-n_2}p^{n_1+n_2}\prescript{}{+1}{Y}_{j_1m_1}(\mathbf{\hat{p}})\prescript{}{-1}{Y}_{j_2m_2}(\mathbf{\hat{p}})\\
{}&\times\frac{1}{\sqrt{4j_1j_2(j_1+1)(j_2+1)}}\Bigg[\Bigg(\left(\frac{\omega j_1(n_1+1)}{p(3-n_1+j_1)}-\frac{pj_1(n_1+3)}{\omega(1-n_1+j_1)}\right)(\vok_{AF}^{(5)})_{n_1j_1m_1}^{(0B)}\\{}&+\frac{5}{n_1+4}(\vok_{AF}^{(5)})^{(1B)}_{n_1j_1m_1}\Bigg)\Bigg(\left(\frac{\omega j_2(n_2+1)}{p(3-n_2+j_2)}-\frac{pj_2(n_2+3)}{\omega(1-n_2+j_2)}\right)(\vok_{AF}^{(5)})_{n_2j_2m_2}^{(0B)}\\{}&+\frac{5}{n_2+4}(\vok_{AF}^{(5)})^{(1B)}_{n_2j_2m_2}\Bigg)+(\vok_{AF}^{(5)})^{(1E)}_{n_1j_1m_1}(\vok_{AF}^{(5)})^{(1E)}_{n_2j_2m_2}\Bigg]\Bigg\}\,.
\end{aligned}
\end{equation}
\end{widetext}

Now that $d$ is fixed, $n$ and $j$ are bounded; hence, the expansion over them can be carried out explicitly:
\begin{widetext}
\begin{equation}
\begin{aligned}
0={}&\left(p_\mu p^\mu\right)^2-4(p_\mu p^\mu)^2\Bigg(\sum_{m}\bigg(\prescript{}{0}{Y}_{0m}(\mathbf{\hat{p}})(\vok_{AF}^{(5)})_{00m}^{(0B)}p+\prescript{}{0}{Y}_{1m}(\mathbf{\hat{p}})\left(\frac{\omega}{3}(\vok_{AF}^{(5)})_{11m}^{(0B)}+\frac{5\omega}{8}(\vok_{AF}^{(5)})_{01m}^{(1B)}\right)\\
{}&+\prescript{}{0}{Y}_{2m}(\mathbf{\hat{p}})\frac{p}{3}(\vok_{AF}^{(5)})_{12m}^{(1B)}\bigg)\Bigg)^2\\
{}&+8p_\mu p^\mu\sum\limits_{m_1m_2}\left(\begin{array}{r l}
(\vok_{AF}^{(5)})^{(0B)}_{11m_1}(\vok_{AF}^{(5)})^{(0B)}_{11m_2} {}&\times\frac{1}{4}\left(\frac{2}{3}\omega^2-4p^2\right)^2\prescript{}{+1}{Y}_{1m_1}(\mathbf{\hat{p}})\prescript{}{-1}{Y}_{1m_2}(\mathbf{\hat{p}})\\{}
+(\vok_{AF}^{(5)})^{(0B)}_{11m_1}(\vok_{AF}^{(5)})^{(1B)}_{01m_2} {}&\times\frac{5}{16}p^2\left(\frac{2}{3}\omega^2-4p^2\right)\prescript{}{+1}{Y}_{1m_1}(\mathbf{\hat{p}})\prescript{}{-1}{Y}_{1m_2}(\mathbf{\hat{p}})\\{}
+(\vok_{AF}^{(5)})^{(0B)}_{11m_2}(\vok_{AF}^{(5)})^{(1B)}_{01m_1} {}&\times\frac{5}{16}p^2\left(\frac{2}{3}\omega^2-4p^2\right)\prescript{}{+1}{Y}_{1m_1}(\mathbf{\hat{p}})\prescript{}{-1}{Y}_{1m_2}(\mathbf{\hat{p}})\\{}
+(\vok_{AF}^{(5)})^{(0B)}_{11m_1}(\vok_{AF}^{(5)})^{(1B)}_{12m_2} {}&\times\frac{1}{4\sqrt{3}}\omega p\left(\frac{2}{3}\omega^2-4p^2\right)\prescript{}{+1}{Y}_{1m_1}(\mathbf{\hat{p}})\prescript{}{-1}{Y}_{2m_2}(\mathbf{\hat{p}})\\{}
+(\vok_{AF}^{(5)})^{(0B)}_{11m_2}(\vok_{AF}^{(5)})^{(1B)}_{12m_1} {}&\times\frac{1}{4\sqrt{3}}\omega p\left(\frac{2}{3}\omega^2-4p^2\right)\prescript{}{+1}{Y}_{1m_1}(\mathbf{\hat{p}})\prescript{}{-1}{Y}_{2m_2}(\mathbf{\hat{p}})\\{}
+(\vok_{AF}^{(5)})^{(1B)}_{01m_1}(\vok_{AF}^{(5)})^{(1B)}_{01m_2} {}&\times\frac{25}{64}p^4\prescript{}{+1}{Y}_{1m_1}(\mathbf{\hat{p}})\prescript{}{-1}{Y}_{1m_2}(\mathbf{\hat{p}})\\{}
+(\vok_{AF}^{(5)})^{(1B)}_{01m_1}(\vok_{AF}^{(5)})^{(1B)}_{12m_2} {}&\times\frac{5}{16\sqrt{3}}p^3\omega\prescript{}{+1}{Y}_{1m_1}(\mathbf{\hat{p}})\prescript{}{-1}{Y}_{2m_2}(\mathbf{\hat{p}})\\{}
+(\vok_{AF}^{(5)})^{(1B)}_{01m_2}(\vok_{AF}^{(5)})^{(1B)}_{12m_1} {}&\times\frac{5}{16\sqrt{3}}p^3\omega\prescript{}{+1}{Y}_{1m_1}(\mathbf{\hat{p}})\prescript{}{-1}{Y}_{2m_2}(\mathbf{\hat{p}})\\{}
+(\vok_{AF}^{(5)})^{(1B)}_{12m_1}(\vok_{AF}^{(5)})^{(1B)}_{12m_2} {}&\times\frac{1}{12}\omega^2p^2\prescript{}{+1}{Y}_{2m_1}(\mathbf{\hat{p}})\prescript{}{-1}{Y}_{2m_2}(\mathbf{\hat{p}})\\{}
+(\vok_{AF}^{(5)})^{(1E)}_{11m_1}(\vok_{AF}^{(5)})^{(1E)}_{11m_2} {}&\times\frac{1}{4}\omega^2p^2\prescript{}{+1}{Y}_{1m_1}(\mathbf{\hat{p}})\prescript{}{-1}{Y}_{1m_2}(\mathbf{\hat{p}})\\{}
+(\vok_{AF}^{(5)})^{(1E)}_{11m_1}(\vok_{AF}^{(5)})^{(1E)}_{22m_2} {}&\times\frac{1}{4\sqrt{3}}\omega p^3\left(\prescript{}{+1}{Y}_{1m_1}(\mathbf{\hat{p}})\prescript{}{-1}{Y}_{2m_2}(\mathbf{\hat{p}})+\prescript{}{+1}{Y}_{2m_1}(\mathbf{\hat{p}})\prescript{}{-1}{Y}_{1m_2}(\mathbf{\hat{p}})\right)\\{}+(\vok_{AF}^{(5)})^{(1E)}_{22m_1}(\vok_{AF}^{(5)})^{(1E)}_{22m_2} {}&\times\frac{1}{12} p^4\prescript{}{+1}{Y}_{2m_1}(\mathbf{\hat{p}})\prescript{}{-1}{Y}_{2m_2}(\mathbf{\hat{p}})
\end{array}\right)\,. \label{vo dispersion relation in component form for d5}
\end{aligned}
\end{equation}
\end{widetext}

This is the most generic dispersion relation for the \emph{leading order vacuum orthogonal model of {\it nm}SME photon sector}. In the isotropic limit, all coefficients except $(\vok_{AF}^{(5)})^{(0B)}_{00m}$ dies, which turns \refe{\ref{vo dispersion relation in component form for d5}} into \refe{\ref{vod in 5}}: a trivial consistency check.

Now that the most generic dispersion relation is obtained, the attention can be restricted to the  interested set $\{\hat{k}_{AF}^{(cn)}, \hat{k}_{AF}^{(bf)}\}$. Then, the last term in the \refe{\ref{vo dispersion relation in component form for d5}} dies out, simplifying the dispersion relation to the form
\begin{equation}
\begin{aligned}
0={}\left(p_\mu p^\mu\right)^2-4(p_\mu p^\mu)^2\Bigg(\sum_{m}\bigg({}&\prescript{}{0}{Y}_{0m}(\mathbf{\hat{p}})(\vok_{AF}^{(5)})_{00m}^{(0B)}p+\prescript{}{0}{Y}_{1m}(\mathbf{\hat{p}})\left(\frac{\omega}{3}(\vok_{AF}^{(5)})_{11m}^{(0B)}+\frac{5\omega}{8}(\vok_{AF}^{(5)})_{01m}^{(1B)}\right)\\{}& +\prescript{}{0}{Y}_{2m}(\mathbf{\hat{p}})\frac{p}{3}(\vok_{AF}^{(5)})_{12m}^{(1B)}\bigg)\Bigg)^2\,.\label{simplified vo dispersion relation in component form for d5}
\end{aligned}
\end{equation}

The equation can be reorganized as
\begin{widetext}
\begin{equation*}
\begin{aligned}
0={}&(p_\mu p^\mu)^2\times\Bigg\{1-2\sum_{m}\Bigg[\prescript{}{0}{Y}_{0m}(\mathbf{\hat{p}})(\vok_{AF}^{(5)})_{00m}^{(0B)}p+\prescript{}{0}{Y}_{1m}(\mathbf{\hat{p}})\left(\frac{\omega}{3}(\vok_{AF}^{(5)})_{11m}^{(0B)}+\frac{5\omega}{8}(\vok_{AF}^{(5)})_{01m}^{(1B)}\right)\\{}&+\prescript{}{0}{Y}_{2m}(\mathbf{\hat{p}})\frac{p}{3}(\vok_{AF}^{(5)})_{12m}^{(1B)}\Bigg]\Bigg\}\Bigg\{1+2\sum_{m}\Bigg[\prescript{}{0}{Y}_{0m}(\mathbf{\hat{p}})(\vok_{AF}^{(5)})_{00m}^{(0B)}p\\
{}&+\prescript{}{0}{Y}_{1m}(\mathbf{\hat{p}})\left(\frac{\omega}{3}(\vok_{AF}^{(5)})_{11m}^{(0B)}+\frac{5\omega}{8}(\vok_{AF}^{(5)})_{01m}^{(1B)}\right)+\prescript{}{0}{Y}_{2m}(\mathbf{\hat{p}})\frac{p}{3}(\vok_{AF}^{(5)})_{12m}^{(1B)}\Bigg]\Bigg\}\,.
\end{aligned}
\end{equation*}
\end{widetext}

At last, it is obvious now to extract the roots:
\begin{equation*}
\begin{aligned}
\omega={}& p\,,\\
\omega={}&\pm\frac{1}{2a}-\frac{b}{a}p
\end{aligned}
\end{equation*}
where
\begin{equation*}
\begin{aligned}
a:={}&\sum\limits_m\prescript{}{0}{Y}_{1m}(\mathbf{\hat{p}})\left(\frac{1}{3}(\vok_{AF}^{(5)})_{11m}^{(0B)}+\frac{5}{8}(\vok_{AF}^{(5)})_{01m}^{(1B)}\right)\,,\\{}
b:={}&\sum\limits_m\bigg(\prescript{}{0}{Y}_{0m}(\mathbf{\hat{p}})(\vok_{AF}^{(5)})_{00m}^{(0B)}+\prescript{}{0}{Y}_{2m}(\mathbf{\hat{p}})\frac{1}{3}(\vok_{AF}^{(5)})_{12m}^{(1B)}\bigg)\,.
\end{aligned}
\end{equation*}

As promised, the nonconventional roots are explicitly spurious. Again, like the result in the isotropic limit, the spurious nature is given by the first term which is expected to remain at all orders, where consideration of higher orders will simply bring higher order perturbative terms into the equation. This is analogous to \refe{3.4} of \refs{\cite{Shreck2013}}, which is the result of a similar calculation within CPT-even sector of {\it nm}SME.

The result is simply that $\hat{k}_{AF}^{(cn)}$ is the only option for the current vacuum orthogonal model, which was deduced at the end of \refp{\ref{s:VODR}} and expressed as vacuum orthogonal model being vacuum orthogonal at all orders. However, the question of whether there indeed exists a nontrivial coefficient subspace\footnote{The trivial coefficient subspace would be the null set, which indicates no Lorentz violation whatsoever in the vacuum orthogonal model of {\it nm}SME photon sector.} which satisfies the necessary conditions in \reft{\ref{Table-vos}} is not addressed.

We explicitly showed that\footnote{See the details in \refa{\ref{App:COD5VOCS}}.} there indeed exists such a nontrivial coefficient subspace, which we shall denote as $\vok_{AF}^{(5cn)}$. In this notation, $AF$ indicates that the coefficient space is CPT-odd, negation diacritic stands for vacuum-orthogonal  model, 5 is the operator dimension and $cn$ denotes that the resultant dispersion relation is conventional. Consequently, $\vok_{AF}^{(5cn)}$ reads as \emph{the coefficient subspace of leading order vacuum orthogonal model of {\it nm}SME CPT-odd modified photon with conventional solutions}.

In this subspace, with \refe{47} of \refs{\cite{Kostelecky2009}}, $(\hat{k}_{AF})_0$ can be shown to take the form
\begin{equation}
(\hat{k}_{AF})_0=p^2\sum\limits_{m}\prescript{}{0}{Y}_{jm}(\mathbf{\hat{p}})(k_{AF}^{(5)})^{(0B)}_{11m}\,,\nonumber
\end{equation}
where contributions of $(\vok_{AF}^{(5)})^{(0B)}_{20m}$ and $(\vok_{AF}^{(5)})^{(0B)}_{00m}$ cancel one another. As $\prescript{}{0}{Y}_{jm}$ are ordinary spherical harmonics, we can rewrite this as
\begin{equation}
(\hat{k}_{AF})_0=p^2\sum\limits_{m}{Y}_{j}^{m}(\mathbf{\hat{p}})(k_{AF}^{(5)})^{(0B)}_{11m}\,,\nonumber
\end{equation}
which then can be inserted into \refe{\ref{eq: Physical vacuum orthogonal propagator}} to yield the corresponding propagator. The important consequence of this is that in addition to the dispersion relation and the polarization vectors, the propagator also remains conventional if the LVT are restricted to $\{(\vok_{AF}^{(5)})^{(0B)}_{000}, (\vok_{AF}^{(5)})^{(0B)}_{200}\}$. The results are summarized in \reft{\ref{Table-cslovom}}.
\begin{table*}
\caption{\label{Table-cslovom}The coefficient subspace of leading order vacuum orthogonal model $\vok_{AF}^{(5c)}$}
\begin{ruledtabular}
\begin{tabular}{l l}
Free Coefficients: & $(\vok_{AF}^{(5)})^{(0B)}_{00m}$ \& $(\vok_{AF}^{(5)})^{(0B)}_{11m}$\\\hline
Nonzero Coefficients:& $(\vok_{AF}^{(5)})^{(0B)}_{00m}$, $(\vok_{AF}^{(5)})^{(0B)}_{11m}$, $(\vok_{AF}^{(5)})^{(0B)}_{20m}$, $(\vok_{AF}^{(5)})^{(1B)}_{01m}$ \& $(\vok_{AF}^{(5)})^{(1B)}_{11m}$\\\hline
Constraint Relations:& $\begin{aligned}
(\vok_{AF}^{(5)})^{(0B)}_{20m}={}&-(\vok_{AF}^{(5)})^{(0B)}_{00m}\,,\\
(\vok_{AF}^{(5)})^{(1B)}_{01m}={}&\frac{296}{109}(\vok_{AF}^{(5)})^{(0B)}_{11m}\,,\\
(\vok_{AF}^{(5)})^{(1B)}_{21m}={}&-\frac{8}{109}(\vok_{AF}^{(5)})^{(0B)}_{11m}
\end{aligned}$\\\hline
Field Theoretical Properties& \begin{tabular}{l}
Conventional Dispersion Relation\\
Conventional Polarization Vectors\\
Conventional Propagator if $(\vok_{AF}^{(5)})^{(0B)}_{11m}=0$
\end{tabular}
\end{tabular}
\end{ruledtabular}
$^*$Parameter $m$ runs from $-j$ to $j$ as integers. 
\end{table*}
\section{Conclusions}
\label{sec:C}
In this study, the CPT-odd {\it nm}SME for the photon sector is analyzed generically. The dispersion relation is calculated for this model out of the general photon sector dispersion relation \cite{Kostelecky2009} and is explicitly expressed in the helicity basis. In this explicit expansion, it is shown that the general dispersion relation can be highly simplified via removal of the complicated terms, once the attention is restricted to $(\hat{k}_{AF})_\pm$. In this LVT regime, coefficient subspaces divide into two, which are named as $\hat{k}_{AF}^{(bf)}$ and $\hat{k}_{AF}^{(cn)}$  where the distinguishing letters stand for the resultant solutions of the relevant coefficient subspace: birefringent, conventional. The remaining coefficient space, which was left beyond the scope of this paper, is denoted as $\hat{k}_{AF}^{(na)}$, where $na$ stands for \emph{Not Analyzed}.

One direct consequence of this result is that there is a possible LVT combination which modifies neither dispersion relation nor polarization vectors for the CPT-odd model. Another important consequence is that there is no coefficient subspace in the analyzed CPT-odd {\it nm}SME model for the photon sector that yields nonconventional nonbirefringent solutions. The second consequence becomes particularly important in the general vacuum orthogonal models. These models are characterized with the fact that they induce no leading order vacuum propagation effect; hence, the initial anticipation would be $\hat{k}_{AF}^{(bf)}$ having no projection on vacuum orthogonal model. A puzzle arises in the model at the first glance, as it was demonstrated that $\hat{k}_{AF}^{(bf)}$ indeed generates solutions in vacuum orthogonal model; however, the resultant solutions are shown to be spurious, which are the solutions that diverge as LV is turned off. It is stated in the \refs{\cite{Kostelecky2009,Shreck2013}} that these solutions are Planck scale effects and should be neglected; therefore, the only possibility with physical solutions for vacuum orthogonal models is $\hat{k}_{AF}^{(cn)}$.

Vacuum orthogonal models are constructed with no leading order vacuum effect; however, our result indicates that vacuum orthogonal models in the CPT-odd sector are vacuum orthogonal at all orders; in other words, they do not accept any solution other than the two standard transverse polarizations with the conventional dispersion relation $\omega=p$ at any order. The only missing part, whether coefficient subspace $\hat{k}_{AF}^{(cn)}$ is nontrivial, is investigated for the leading order model, and is explicitly shown to be nontrivial: There arises a two parameter coefficient subspace, denoted by $\vok_{AF}^{(5c)}$, which induces no effect at all orders, both on polarization vectors and on the dispersion relations in the leading order model.

In addition to the dispersion relations and the polarization vectors, the general propagator is also addressed for the {\it nm}SME CPT-odd modified photon sector. As the generic form of inverse propagator is an infinite series, it was argued that it would be formidable, if possible, to invert it covariantly and analytically; hence, the propagator is calculated explicitly in the Helicity basis, which was able to give an analytically exact form, albeit covariance is lost.

The propagator is also calculated for the vacuum orthogonal special case and shown to be non-vanishing unless all $(\hat{k}_{AF})_i$ are explicitly zero. Additionally, it is demonstrated all $(\hat{k}_{AF})_i$ vanish if all Lorentz violation is provided with two non-vanishing terms $(\vok_{AF}^{(5)})^{(0B)}_{000}$ \& $(\vok_{AF}^{(5)})^{(0B)}_{200}$ with $(\vok_{AF}^{(5)})^{(0B)}_{200}=-(\vok_{AF}^{(5)})^{(0B)}_{000}$.

\appendix
\addtocontents{toc}{\protect\setcounter{tocdepth}{0}}	
\section{General Dispersion Relation in the Helicity Basis}
\label{App:HB}
\subsection{Helicity Basis}
The helicity basis is a three dimensional basis, which exploits the relations between angular momentum, helicity and spin-weighted spherical harmonics in a convenient way. As mentioned in the introduction, the spherical decomposition is the relevant decomposition for observations and experiments, and hence were introduced in the photon sector in \refs{\cite{Kostelecky2009}}. Although the details regarding the helicity basis are available in the Appendix A of last reference, we will provide a quick summary here.

In a nutshell, helicity basis is \emph{complex spherical polar coordinates}, and the usual spherical polar angles $\theta$ and $\phi$ mix to result in positive and negative helicities
\begin{equation*}
\hat{\bm{e}}_\pm=\hat{\bm{e}}^\mp=\frac{1}{\sqrt{2}}\left(\hat{\bm{e}}_\theta\pm i\hat{\bm{e}}_\phi\right)\,,
\end{equation*}
where the radial direction remains as it is: $\hat{\bm{e}}_r=\hat{\bm{e}}^r=\hat{\bm{p}}$.
In helicity basis, the Metric and the Levi-Civita take the form
\begin{equation}
\begin{aligned}
&{}g_{ab}=g^{ab}=\begin{pmatrix}
0& 0& 1\\ 0& 1& 0\\ 1& 0& 0
\end{pmatrix}\,,\\
&{}\epsilon_{+r-}=-\epsilon^{+r-}=i\,,
\end{aligned}\label{Equation: Helicity-levi-civita}
\end{equation} 
where positive signature is used as spacepart of the full metric is dealt here only. That means, the full metric whose space part is in helicity basis is given as
\begin{equation}
\eta^{\mu\nu}=\eta_{\mu\nu}=\begin{pmatrix}
1& 0& 0& 0\\
0& 0& 0& -1\\
0& 0& -1& 0\\
0& -1& 0& 0
\end{pmatrix}\,.\label{Equation: full helicity metric}
\end{equation}

\subsection{Dispersion Relation}
The most general form of the dispersion relation for the {\it nm}SME photon sector is given by \refe{30} of \refs{\cite{Kostelecky2009}}, which reads as
\begin{equation*}
0=-\frac{1}{3}\epsilon_{\mu_1\mu_2\mu_3\mu_4}\epsilon_{\nu_1\nu_2\nu_3\nu_4}p_{\rho_1}p_{\rho_2}p_{\rho_3}p_{\rho_4}\hat{\chi}^{\mu_1\mu_2\nu_1\rho_1}\hat{\chi}^{\nu_2\rho_2\rho_3\mu_3}\hat{\chi}^{\rho_4\mu_4\nu_3\nu_4}+8p_\alpha p_\beta(\hat{k}_{AF})_\mu(\hat{k}_{AF})_\nu\hat{\chi}^{\alpha\mu\beta\nu}\,,
\end{equation*}
where the 4-tensor $\hat{\chi}^{\mu\nu\rho\sigma}$ is defined in the same reference as
\begin{equation*}
\hat{\chi}^{\mu\nu\rho\sigma}=\frac{1}{2}\left(\eta^{\mu\rho}\eta^{\nu\sigma}-\eta^{\nu\rho}\eta^{\mu\sigma}\right)+(\hat{k}_F)^{\mu\nu\rho\sigma}\,.
\end{equation*}

Once $(\hat{k}_F)^{\mu\nu\rho\sigma}=0$ is imposed and $\hat{\chi}$ are inserted, the equation can be expanded term by term. In each term, one of the Levi-Civita tensor can be raised to the contravariant form via the formula
\begin{equation*}
\epsilon^{\mu_1\nu_1\rho_1\sigma_1}=-\eta^{\mu_1\mu_2}\eta^{\nu_1\nu_2}\eta^{\rho_1\rho_2}\eta^{\sigma_1\sigma_2}\epsilon_{\mu_2\nu_2\rho_2\sigma_2}\,,
\end{equation*}
where the minus sign is required as the four-dimensional spacetime is of Minkowskian signature. Then, with the contractions of the Levi-Civita tensors, the dispersion relation reduces to \refe{\ref{General CPT-odd Dispersion Relation}}
\begin{equation*}
0=\big(p_\mu p^\mu\big)^2+4p_\alpha p^\alpha(\hat{k}_{AF})_\mu(\hat{k}_{AF})^\mu-4\big( p_\mu(\hat{k}_{AF})^\mu\big)^2\,.
\end{equation*}

In order to spherically decompose this equation, one only needs to expand the contractions via
the metric \refe{\ref{Equation: full helicity metric}}, with which
\begin{equation*}
\begin{aligned}
p_\mu(\hat{k}_{AF})^\mu=&{}\omega(\hat{k}_{AF})_0-p(\hat{k}_{AF})_r\,,\\
(\hat{k}_{AF})_\mu(\hat{k}_{AF})^\mu=&{}\left((\hat{k}_{AF})_0\right)^2-\left((\hat{k}_{AF})_r\right)^2-2(\hat{k}_{AF})_+(\hat{k}_{AF})_-\,,
\end{aligned}
\end{equation*}
for usual frequency $\omega$ and spacepart magnitude $p$. With these substitution, \refe{\ref{General CPT-odd Dispersion Relation}} reduces to \refe{\ref{General CPT-odd Spherical Dispersion Relation}}.

A consistency check of \refe{\ref{General CPT-odd Dispersion Relation}} can be conducted for the vacuum model. The leading order condition ($\omega\simeq p$) for this model gets second term neglected, hence
\begin{equation}
0\simeq\big(p_\mu p^\mu\big)^2-4\big( p_\mu(\hat{k}_{AF})^\mu\big)^2\,.\label{Eq:App:1}
\end{equation}
The general dispersion relation for the vacuum model is given by \refe{74} of \refs{\cite{Kostelecky2009}} as
\begin{equation}
(p_\mu p^\mu-(\hat{c}_F)^{\mu\nu}p_\mu p_\nu)^2-2(\hat{\chi}_\omega)^{\alpha\beta\gamma\delta}(\hat{\chi}_\omega)_{\alpha\mu\gamma\nu}p_\beta p_\delta p^\mu p^\nu -4(p^\mu(\hat{k}_{AF})_\mu)^2\simeq 0\,,\label{Eq:App:2}
\end{equation}
where $(\hat{\chi}_\omega)_{\alpha\mu\gamma\nu}$ denotes the Weyl component of constitutive tensor $\hat{\chi}$.

The condition $\hat{k}_F=0$ for special CPT-odd model kills both $\hat{c}_F$ and $\hat{\chi}_\omega$, as can be deduced from \refe{40} of \refs{\cite{Kostelecky2009}}, hence reduces \refe{\ref{Eq:App:2}} to \refe{\ref{Eq:App:1}}, which completes the consistency check.

\section{Calculation of Vacuum Orthogonal Dispersion Relation}
\label{App:Covodr}
The prescription \refe{97-98} of \refs{\cite{Kostelecky2009}} in order to restrict to the vacuum orthogonal model can be applied straightforwardly to the general CPT-odd {\it nm}SME photon sector dispersion relation  (\ref{General CPT-odd Dispersion Relation2}):
\begin{widetext}
\begin{equation*}
\begin{aligned}
0=&\left(p_\mu p^\mu\right)^2-4\Bigg\{\sum\limits_{dnjm}\omega^{d-3-n}p^n\prescript{}{0}{Y}_{jm}(\mathbf{\hat{p}})\Bigg[\frac{dp}{n+3}\bigg(\frac{(d-2-n)(n+3)}{d(d-2-n+j)}\left((\vok_{AF}^{(d)})^{(0B)}_{njm}-(\vok_{AF}^{(d)})^{(0B)}_{(n-2)jm}\right)\\&-\frac{1}{n+1}(\vok_{AF}^{(d)})^{(1B)}_{(n-1)jm}\bigg)+\frac{\omega}{n+2}\bigg(\frac{j(n+2)}{d-3-n+j}\left((\vok_{AF}^{(d)})^{(0B)}_{(n+1)jm}-(\vok_{AF}^{(d)})^{(0B)}_{(n-1)jm}\right)\\&+\frac{d}{n+4}(\vok_{AF}^{(d)})^{(1B)}_{njm}\bigg)\Bigg]\Bigg\}^2\\&-8p_\mu p^\mu\sum\limits_{d_1d_2n_1n_2j_1j_2m_1m_2}\omega^{d_1+d_2-6-n_1-n_2}p^{n_1+n_2}\prescript{}{+1}{Y}_{j_1m_1}(\mathbf{\hat{p}})\prescript{}{-1}{Y}_{j_2m_2}(\mathbf{\hat{p}})\frac{1}{\sqrt{4j_1j_2(j_1+1)(j_2+1)}}\\
&\times\Bigg(\left(\frac{j_1(n_1+2)}{d_1-3-n_1+j_1}\left((\vok_{AF}^{(d_1)})^{(0B)}_{(n_1+1)j_1m_1}-(\vok_{AF}^{(d_1)})^{(0B)}_{(n_1-1)j_1m_1}\right)+\frac{d_1}{n_1+4}(\vok_{AF}^{(d_1)})^{(1B)}_{n_1j_1m_1}\right)+i(\vok_{AF}^{(d_1)})^{(1E)}_{n_1j_1m_1}\Bigg)\\&\times\Bigg(\left(\frac{-j_2(n_2+2)}{d_2-3-n_2+j_2}\left((\vok_{AF}^{(d_2)})^{(0B)}_{(n_2+1)j_2m_2}-(\vok_{AF}^{(d_2)})^{(0B)}_{(n_2-1)j_2m_2}\right)-\frac{d_2}{n_2+4}(\vok_{AF}^{(d_2)})^{(1B)}_{n_2j_2m_2}\right)+i(\vok_{AF}^{(d_2)})^{(1E)}_{n_2j_2m_2}\Bigg)\,.
\end{aligned}
\end{equation*}
\end{widetext}

As the last two rows are of the form $\left(A(t_1)+iB(t_1)\right)\left(-A(t_2)+iB(t_2)\right)$ where $t_i$ is the collective index for $\{d_i,n_i,j_i,m_i\}$, the imaginary part of the overall term is antisymmetric over $\{t_1,t_2\}$, which dies in the summation which is symmetric over these indices. Therefore, the dispersion relation reduces to the form:
\begin{equation}
0=(p_\mu p^\mu)^2-\mathcal{T}_1(\omega,p)+(p_\mu p^\mu)\mathcal{Q}(\omega,p)\,, \label{dispersion relation in terms of T}
\end{equation}
where
\begin{equation}
\begin{aligned}
\mathcal{Q}(\omega,p):=& 8\sum\limits_{d_1d_2n_1n_2j_1j_2m_1m_2}\omega^{d_1+d_2-6-n_1-n_2}p^{n_1+n_2}\prescript{}{+1}{Y}_{j_1m_1}(\mathbf{\hat{p}})\prescript{}{-1}{Y}_{j_2m_2}(\mathbf{\hat{p}})\frac{1}{\sqrt{4j_1j_2(j_1+1)(j_2+1)}}\\
&\times\Bigg\{\Bigg(\left(\frac{\omega j_1(n_1+1)}{p(d_1-2-n_1+j_1)}-\frac{pj_1(n_1+3)}{\omega(d_1-4-n_1+j_1)}\right)(\vok_{AF}^{(d_1)})_{n_1j_1m_1}^{(0B)}+\frac{d_1}{n_1+4}(\vok_{AF}^{(d_1)})^{(1B)}_{n_1j_1m_1}\Bigg)\\&\times\Bigg(\left(\frac{\omega j_2(n_2+1)}{p(d_2-2-n_2+j_2)}-\frac{pj_2(n_2+3)}{\omega(d_2-4-n_2+j_2)}\right)(\vok_{AF}^{(d_2)})_{n_2j_2m_2}^{(0B)}+\frac{d_2}{n_2+4}(\vok_{AF}^{(d_2)})^{(1B)}_{n_2j_2m_2}\Bigg)\\&+(\vok_{AF}^{(d_1)})^{(1E)}_{n_1j_1m_1}(\vok_{AF}^{(d_2)})^{(1E)}_{n_2j_2m_2}\Bigg\}\,,
\end{aligned}\label{definition of Q}
\end{equation}
and where
\begin{equation*}
\begin{aligned}
\mathcal{T}_1(\omega,p):=& 4\Bigg\{\sum\limits_{dnjm}\omega^{d-3-n}p^n\prescript{}{0}{Y}_{jm}(\mathbf{\hat{p}})\bigg[(\vok_{AF}^{(d)})^{(0B)}_{njm}\frac{p}{\omega^2}\left(\frac{(d-2-n)\omega^2}{d-2-n+j}-\frac{(d-4-n)p^2}{d-4-n+j}\right)\\&-\frac{dp^2}{(n+4)(n+2)\omega}(\vok_{AF}^{(d)})^{(1B)}_{njm}+(\vok_{AF}^{(d)})^{(0B)}_{njm}\frac{j}{p}\left(\frac{\omega^2}{d-2-n+j}-\frac{p^2}{d-4-n+j}\right)\\&+\frac{d\omega}{(n+2)(n+4)}(\vok_{AF}^{(d)})^{(1B)}_{njm}\bigg]\Bigg\}^2\,,
\end{aligned}
\end{equation*}
which can be reduced to the form
\begin{equation}
\begin{aligned}
\mathcal{T}_1(\omega,p)=& 4\Bigg\{\sum\limits_{dnjm}\omega^{d-4-n}p^n\prescript{}{0}{Y}_{jm}(\mathbf{\hat{p}})\Bigg((p_\mu p^\mu)(\vok_{AF}^{(d)})^{(1B)}_{njm}\frac{d}{(n+2)(n+4)}\Bigg)+\omega^{d-5-n}p^{n-1}\prescript{}{0}{Y}_{jm}(\mathbf{\hat{p}})\mathcal{T}_2\Bigg\}^2
\end{aligned}\label{equation of T}
\end{equation}
for
\begin{equation*}
\mathcal{T}_2:=(\vok_{AF}^{(d)})^{(0B)}_{njm}\left(p^2\left(\frac{(d-2-n)\omega^2}{d-2-n+j}-\frac{(d-4-n)p^2}{d-4-n+j}\right)+j\omega^2\left(\frac{\omega^2}{d-2-n+j}-\frac{p^2}{d-4-n+j}\right)\right)\,.
\end{equation*}

The range of the coefficients can be deduced from \reft{XVIII} of \refs{\cite{Kostelecky2009}} as in \reft{\ref{Table-irfvoc}}. By algebraic simplifications then, $\mathcal{T}_2$ can be brought to the form
\begin{table*}
\caption{\label{Table-irfvoc}Index ranges for Vacuum Orthogonal Coefficients.}
\begin{ruledtabular}
\begin{tabular}{l@{\hskip 0.3in}c@{\hskip 0.3in}c@{\hskip 0.3in}c@{\hskip 0.3in}}
Coefficient & d & n & j\\ \hline
$(\vok_{AF}^{(d)})^{(0B)}_{njm}$& odd$\ge5$&$0,1,...,d-4$&$n,n-2,n-4,...,\ge0$\\
$(\vok_{AF}^{(d)})^{(1B)}_{njm}$& odd$\ge5$&$0,1,...,d-4$&$n+1,n-1,n-3,...,\ge1$\\
$(\vok_{AF}^{(d)})^{(1E)}_{njm}$& odd$\ge5$&$1,2,...,d-3$&$n,n-2,n-4,...,\ge1$
\end{tabular}
\end{ruledtabular}
\end{table*}
\begin{equation*}
\mathcal{T}_2=(p_\mu p^\mu)(\vok_{AF}^{(d)})^{(0B)}_{njm}\left(\omega^2\frac{j}{d-2-n+j}+p^2\frac{d-4-n}{d-4-n+j}\right)\,,
\end{equation*}
which can be inserted into \refe{\ref{equation of T}}, which is itself to be inserted into \refe{\ref{dispersion relation in terms of T}}. Therefore, the general dispersion relation for the vacuum-orthogonal photon sector of {\it nm}SME becomes
\begin{equation}
0=(p_\mu p^\mu)\times\Big((p_\mu p^\mu)\mathcal{P}(\omega,p)+\mathcal{Q}(\omega,p)\Big)\,,
\end{equation}
where
\begin{equation}
\begin{aligned}
\mathcal{P}(\omega,p):=& 1-4\Bigg\{\sum\limits_{dnjm}\omega^{d-4-n}p^n\prescript{}{0}{Y}_{jm}(\mathbf{\hat{p}})\Bigg((\vok_{AF}^{(d)})^{(1B)}_{njm}\frac{d}{(n+2)(n+4)}\Bigg)\\&+\omega^{d-5-n}p^{n-1}\prescript{}{0}{Y}_{jm}(\mathbf{\hat{p}})(\vok_{AF}^{(d)})^{(0B)}_{njm}\left(\omega^2\frac{j}{d-2-n+j}+p^2\frac{d-4-n}{d-4-n+j}\right)\Bigg\}^2\,,
\end{aligned}
\end{equation}
and  where $\mathcal{Q}(\omega,p)$ is given by \refe{\ref{definition of Q}}.

\section{Calculation of Vacuum Orthogonal \& Isotropic Dispersion Relation in Ring Coefficients}
\label{App:COVOIDRIRC}
The \emph{ring-coefficients} in general are defined by \refe{71} of \refs{\cite{Kostelecky2009}} as
\begin{equation*}
\begin{aligned}
(\ring{c}_{F}^{(d)})_n=&{}(c_F^{(d)})^{(0E)}_{n00}\,,\\
(\ring{k}_{F}^{(d)})_n=&{}(k_F^{(d)})^{(0E)}_{n00}\,,\\
(\ring{k}_{AF}^{(d)})_n=&{}(k_{AF}^{(d)})^{(0B)}_{n00}\,.\\
\end{aligned}
\end{equation*}
According to \refe{97} of \refs{\cite{Kostelecky2009}}, $(k_{AF}^{(d)})^{(0B)}_{n00}=\frac{n+3}{d}\left((\vok_{AF}^{(d)})^{(0B)}_{n00}-(\vok_{AF}^{(d)})^{(0B)}_{(n-2)00}\right)$, hence
\begin{equation}
(\ring{k}_{AF}^{(d)})_n=\frac{n+3}{d}\left((\vok_{AF}^{(d)})^{(0B)}_{n00}-(\vok_{AF}^{(d)})^{(0B)}_{(n-2)00}\right)\label{ring coefficients}\,.
\end{equation}

This equality suggests that it is not straightforward to explicitly write $(\vok_{AF}^{(d)})^{(0B)}_{n00}$ in terms of $(\ring{k}_{AF}^{(d)})_n$, which would allow derivation of dispersion relation in ring coefficients out of \refe{\ref{vacuum orthogonal & isotropic dispersion relation}} in one step. Instead, we start from the most general formula, \refe{\ref{General CPT-odd Spherical Dispersion Relation}}.

In isotropic limit $(\hat{k}_{AF})_\pm=0$; and, $p(\hat{k}_{AF})_0-\omega(\hat{k}_{AF})_r$ becomes
\begin{equation*}
p(\hat{k}_{AF})_0-\omega(\hat{k}_{AF})_r=\frac{1}{\sqrt{4\pi}}\sum\limits_{d=\text{odd}}\sum\limits_{n}\omega^{d-3-n}p^{n+1}\left((\vok_{AF}^{(d)})^{(0B)}_{n00}-(\vok_{AF}^{(d)})^{(0B)}_{(n-2)00}\right) \label{pk0-omegakr}
\end{equation*}
from \refe{47,48,97} of \refs{\cite{Kostelecky2009}}. Therefore, the dispersion relation is
\begin{equation*}
0=(p_\mu p^\mu)^2-\frac{1}{\pi}\left(\sum\limits_{d=\text{odd}}\sum\limits_{n}\omega^{d-3-n}p^{n+1}\left((\vok_{AF}^{(d)})^{(0B)}_{n00}-(\vok_{AF}^{(d)})^{(0B)}_{(n-2)00}\right)\right)^2\,.
\end{equation*}
According to \reft{\ref{Table-irfvoc}}, $n$ takes positive even integer values upto $d-5$ for $j=0$, hence the dispersion relation can be written as
\begin{equation*}
0=(p_\mu p^\mu)^2-\frac{1}{\pi}\left(\sum\limits_{d=\text{odd}}\left(\omega^{d-3}p(\vok_{AF}^{(d)})^{(0B)}_{000}+\sum\limits_{n=2}^{d-5}\omega^{d-3-n}p^{n+1}\frac{d}{n+3}(\ring{k}_{AF}^{(d)})_n-p^{d-2}(\vok_{AF}^{(d)})^{(0B)}_{(d-5)00}\right)\right)^2\,.
\end{equation*}

From Equation \refe{\ref{ring coefficients}}, we can write
\begin{equation*}
(\vok_{AF}^{(d)})^{(0B)}_{(d-5)00}=(\vok_{AF}^{(d)})^{(0B)}_{000}+\sum\limits_{n=2}^{d-5}\frac{d}{n+3}(\ring{k}_{AF}^{(d)})_n\,,
\end{equation*}
then
\begin{equation*}
0=(p_\mu p^\mu)^2-\frac{1}{\pi}\left(\sum\limits_{d=\text{odd}}\sum\limits_{n=0}^{d-5}\frac{d}{n+3}\left(\omega^{d-3-n}-p^{d-3-n}\right)\left((\ring{k}_{AF}^{(d)})_np^{n+1}\right)\right)^2\,.
\end{equation*}

Since $d-3-n$ is always greater or equal to 2, and is always even, we can use the following equality:
\begin{equation*}
\omega^{d-3-n}-p^{d-3-n}=(\omega^2-p^2)\sum\limits_{i=0}^{(d-5-n)/2}\omega^{d-5-n-2i}p^{2i}\,,
\end{equation*}
which in turn gives the final form of dispersion relation in terms of ring coefficients \refe{\ref{disperion relation in ring coefficients}}.

\section{Calculation of $d=5$ Vacuum Orthogonal Coefficient Subspace}
\label{App:COD5VOCS}
For d=5, \refe{\ref{first condition for vacuum polarization}, \ref{second condition for vacuum polarization}} become:
\begin{equation*}
\begin{aligned}
0=&\sum\limits_{n}(\vok_{AF}^{(5)})^{(0B)}_{njm}\left(-\frac{4}{5}+\frac{4j(6+j)}{5(3-n+j)(1-n+j)}\right)\\
&+\sum\limits_{n}(\vok_{AF}^{(5)})^{(1B)}_{njm}\left(\frac{1}{n+2}-\frac{5}{(n+4)(n+2)}-\frac{(2-n)(n+4)}{5(2-n+j)(n+2)}\right)\,,\\
0=&\sum\limits_{n}\left(-\frac{2j(4+j)}{(3-n+j)(1-n+j)}(\vok_{AF}^{(5)})^{(0B)}_{njm}+\frac{5}{n+4}(\vok_{AF}^{(5)})^{(1B)}_{njm}\right)\,,\\
0=&\sum\limits_{n}(\vok_{AF}^{(5)})^{(1E)}_{njm}\,,
\end{aligned}\label{d=5 conditions}
\end{equation*}
where index ranges become as in \reft{\ref{Table-irflovoc}}.
\begin{table*}
\caption{\label{Table-irflovoc}Index ranges for Leading Order Vacuum Orthogonal Coefficients.}
\begin{ruledtabular}
\begin{tabular}{l@{\hskip 0.3in}c@{\hskip 0.3in}c@{\hskip 0.3in}}
Coefficient & n & j\\ \hline
$(\vok_{AF}^{(d)})^{(0B)}_{njm}$&$0,1,2$&$n,n-2\ge0$\\
$(\vok_{AF}^{(d)})^{(1B)}_{njm}$&$0,1,2$&$n+1,n-1\ge1$\\
$(\vok_{AF}^{(d)})^{(1E)}_{njm}$&$1,2$&$n$
\end{tabular}
\end{ruledtabular}
\end{table*}
As these equations are to hold at all possible $j$, one needs to check for each $j$ value separately.
\paragraph{For $j=0$}:
\begin{equation*}
\begin{aligned}
0=&-\frac{4}{5}\sum\limits_{n}(\vok_{AF}^{(5)})^{(0B)}_{n0m}+\sum\limits_{n}(\vok_{AF}^{(5)})^{(1B)}_{n0m}\left(\frac{1}{n+2}-\frac{5}{(n+4)(n+2)}-\frac{(n+4)}{5(n+2)}\right)\,,\\
0=&\sum\limits_{n}\frac{5}{n+4}(\vok_{AF}^{(5)})^{(1B)}_{n0m}\,,\\
0=&\sum\limits_{n}(\vok_{AF}^{(5)})^{(1E)}_{n0m}\,.
\end{aligned}
\end{equation*}
As the only coefficients with $j=0$ are $(\vok_{AF}^{(5)})^{(0B)}_{00m}$ and $(\vok_{AF}^{(5)})^{(0B)}_{20m}$,
\begin{equation}
0=(\vok_{AF}^{(5)})^{(0B)}_{00m}+(\vok_{AF}^{(5)})^{(0B)}_{20m}\,. \label{d=5, j=0 conditions}
\end{equation}
\paragraph{For $j=1$}:
\begin{equation*}
\begin{aligned}
0=&\sum\limits_{n}(\vok_{AF}^{(5)})^{(0B)}_{n1m}\left(-\frac{4}{5}+\frac{28}{5(4-n)(2-n)}\right)\nonumber\\
&+\sum\limits_{n}(\vok_{AF}^{(5)})^{(1B)}_{n1m}\left(\frac{1}{n+2}-\frac{5}{(n+4)(n+2)}-\frac{(2-n)(n+4)}{5(3-n)(n+2)}\right)\,,\\
0=&\sum\limits_{n}\left(-\frac{10}{(4-n)(2-n)}(\vok_{AF}^{(5)})^{(0B)}_{n1m}+\frac{5}{n+4}(\vok_{AF}^{(5)})^{(1B)}_{n1m}\right)\,,\\
0=&\sum\limits_{n}(\vok_{AF}^{(5)})^{(1E)}_{n1m}\,.
\end{aligned}
\end{equation*}
As the only coefficients with $j=1$ are $(\vok_{AF}^{(5)})^{(0B)}_{11m}$, $(\vok_{AF}^{(5)})^{(1B)}_{01m}$, $(\vok_{AF}^{(5)})^{(1B)}_{21m}$ and $(\vok_{AF}^{(5)})^{(1E)}_{11m}$,
\begin{equation}
\begin{aligned}
0=&\frac{16}{15}(\vok_{AF}^{(5)})^{(0B)}_{11m}-\frac{47}{120}(\vok_{AF}^{(5)})^{(1B)}_{01m}+\frac{1}{24}(\vok_{AF}^{(5)})^{(1B)}_{21m}\,,\\
0=&-\frac{10}{3}(\vok_{AF}^{(5)})^{(0B)}_{11m}+\frac{5}{4}(\vok_{AF}^{(5)})^{(1B)}_{01m}+\frac{5}{6}(\vok_{AF}^{(5)})^{(1B)}_{21m}\,,\\
0=&(\vok_{AF}^{(5)})^{(1E)}_{11m}\,.
\end{aligned}\label{d=5, j=1 conditions}
\end{equation}
\paragraph{For $j=2$}:
\begin{equation*}
\begin{aligned}
0=&\sum\limits_{n}(\vok_{AF}^{(5)})^{(0B)}_{n2m}\left(-\frac{4}{5}+\frac{64}{5(5-n)(3-n)}\right)\nonumber\\
&+\sum\limits_{n}(\vok_{AF}^{(5)})^{(1B)}_{n2m}\left(\frac{1}{n+2}-\frac{5}{(n+4)(n+2)}-\frac{(2-n)(n+4)}{5(4-n)(n+2)}\right)\,,\\
0=&\sum\limits_{n}\left(-\frac{24}{(5-n)(3-n)}(\vok_{AF}^{(5)})^{(0B)}_{n2m}+\frac{5}{n+4}(\vok_{AF}^{(5)})^{(1B)}_{n2m}\right)\,,\\
0=&\sum\limits_{n}(\vok_{AF}^{(5)})^{(1E)}_{n2m}\,.
\end{aligned}
\end{equation*}
As the only coefficients with $j=2$ are $(\vok_{AF}^{(5)})^{(0B)}_{22m}$, $(\vok_{AF}^{(5)})^{(1B)}_{12m}$ and $(\vok_{AF}^{(5)})^{(1E)}_{22m}$,
\begin{equation}
\begin{aligned}
0=&\frac{52}{15}(\vok_{AF}^{(5)})^{(0B)}_{22m}-\frac{1}{9}(\vok_{AF}^{(5)})^{(1B)}_{12m}\,,\\
0=&-8(\vok_{AF}^{(5)})^{(0B)}_{22m}+(\vok_{AF}^{(5)})^{(1B)}_{12m}\,,\\
0=&(\vok_{AF}^{(5)})^{(1E)}_{22m}\,.
\end{aligned}\label{d=5, j=2 conditions}
\end{equation}
\paragraph{For $j=3$}:
The only coefficient with $j=3$ is $(\vok_{AF}^{(5)})^{(1B)}_{23m}$. Hence,
\begin{equation}
\begin{aligned}
0=&\frac{1}{24}(\vok_{AF}^{(5)})^{(1B)}_{23m}\,,\\
0=&\frac{5}{6}(\vok_{AF}^{(5)})^{(1B)}_{23m}\,.	
\end{aligned}\label{d=5, j=3 conditions}
\end{equation}

From \refe{\ref{d=5, j=0 conditions}, \ref{d=5, j=1 conditions}, \ref{d=5, j=2 conditions}, \ref{d=5, j=3 conditions}}, one arrives at the results summarized in \reft{\ref{Table-cslovom}}.

\begin{acknowledgements}
I.T. and S.A. thank  METU-BAP grant number 08-11-2013-028. S.A. also thanks T\"{U}B\.ITAK for its partial support. I.T. also thanks TUBA-GEBIP for its partial support. We are grateful to Alan Kostelecky for his help and fruitful email correspondences throughout this study.
\end{acknowledgements}



\begin{thebibliography}{99}

\bibitem{String_Theory}
V.A.~Kosteleck\'y and S.~Samuel, Phys. Rev. D 39, 683 (1989); V.A.~Kosteleck\'y and R.~Potting, Nucl. Phys. B 359, 545 (1991); V.A.~Kosteleck\'y and R.~Potting, Phys. Rev. D 51, 3923 (1995), hep-ph/9501341.
\bibitem{Gambini}
R.~Gambini and J.~Pullin, Phys. Rev. D 59, 124021 (1999), gr-qc/9809038.
\bibitem{Loop QG}
M.~Bojowald, H.A.~Morales-Técotl, and H.~Sahlmann, Phys. Rev. D 71, 084012 (2005), gr-qc/0411101.
\bibitem{Non-commutative G}
S.M.~Carroll, J.A.~Harvey, V.A.~Kosteleck\'y, C.~D.~Lane, and T.~Okamoto, Phys. Rev. Lett. 87, 141601 (2001), hep-th/0105082.
\bibitem{LV Theories}
J.A.~Wheeler, Ann. Phys. (N.Y.) 2, 604 (1957); S.W.~Hawking, Nucl. Phys. B 144, 349 (1978); F.R.~Klinkhamer, Nucl. Phys. B 535, 233 (1998), hep-th/9805095; F.R.~Klinkhamer, Nucl. Phys. B 578, 277 (2000), hep-th/9912169; F.R.~Klinkhamer and C.~Rupp, Phys. Rev. D 70, 045020 (2004), hep-th/0312032; S.~Bernadotte and F.R.~Klinkhamer, Phys. Rev. D 75, 024028 (2007), hep-ph/0610216.

\bibitem{Kinematical LV}
H.P.~Robertson, Rev. Mod. Phys. 21, 378 (1949); R.~Mansouri and R.U.~Sexl, Gen. Relativ. Gravit. 8, 497 (1977).
\bibitem{Field T. LV}
D.~Sudarsky, L.~Urrutia, and H.~Vucetich, Phys. Rev. Lett. 89, 231301 (2002); J.~Alfaro, H.A.~Morales-Técotl, and L.F.~Urrutia, Phys. Rev. D 65, 103509 (2002); R.C.~Myers and M.~Pospelov, Phys. Rev. Lett. 90, 211601 (2003); A.G.~Cohen and S.L.~Glashow, Phys. Rev. Lett. 97, 021601 (2006); C.M.~Reyes, L.F.~Urrutia, and J.D.~Vergara, Phys. Rev. D 78, 125011 (2008); P.A.~Bolokhov and M.~Pospelov, Phys. Rev. D 77, 025022 (2008).

\bibitem{Fermion}
V.A.~Kosteleck\'y, M.~Mewes, Phys. Rev. D 88, 096006 (2013), arXiv:1308.4973 [hep-ph].

\bibitem{Kostelecky2009}
V.A.~Kosteleck\'y and M.~Mewes, Phys. Rev. D 80, 015020 (2009), arXiv:0905.0031 [hep-ph].
\bibitem{Cambiaso:2012vb} 
  M.~Cambiaso, R.~Lehnert and R.~Potting,
  Phys.\ Rev.\ D {\bf 85}, 085023 (2012)
  [arXiv:1201.3045 [hep-th]].
  
\bibitem{Gravity}
V.A.~Kosteleck\'y, Phys. Rev. D 69, 105009 (2004), hep-th/0312310v2.

\bibitem{Kostelecky:2011qz} 
A.~Kostelecky,
Phys.\ Lett.\ B {\bf 701}, 137 (2011)
[arXiv:1104.5488 [hep-th]].


\bibitem{SME}
D.~Colladay and V.A.~Kosteleck\'y, Phys. Rev. D 55, 6760 (1997), hep-ph/9703464.

\bibitem{Colladay.Kostelecky}
D.~Colladay and V.A.~Kosteleck\'y, Phys. Rev. D 58, 116002 (1998), hep-ph/9809521.

\bibitem{Kostelecky:2008ts} 
V.~A.~Kostelecky and N.~Russell,
Rev.\ Mod.\ Phys.\  {\bf 83}, 11 (2011)
[arXiv:0801.0287 [hep-ph]].

\bibitem{Neutrino}
V.A.~Kosteleck\'y and M.~Mewes, Phys. Rev. D 85, 096005 (2012), arXiv:1112.6395 [hep-ph].

\bibitem{Kostelecky:2008tsupdates} 
The updates of the tables can be followed in the arXiv entry, arXiv:0801.0287 [hep-ph].


\bibitem{EFT}
H.~Georgi, Ann. Rev. Nucl. Part. Sci. 43, 209 (1993); A.~Pich, hep-ph/9806303.


\bibitem{Kostelecky:2001mb} 
V.~A.~Kostelecky and M.~Mewes,
Phys.\ Rev.\ Lett.\  {\bf 87}, 251304 (2001)
[hep-ph/0111026];
V.~A.~Kostelecky and M.~Mewes,
Phys.\ Rev.\ D {\bf 66}, 056005 (2002)
[hep-ph/0205211].

\bibitem{Shreck2013}
M.~Schreck, Phys. Rev. D 89, 105019 (2014), arXiv:1312.4916 [hep-th].

\bibitem{Schreck:2013gma} 
M.~Schreck,
Phys.\ Rev.\ D {\bf 89}, no. 8, 085013 (2014)
[arXiv:1311.0032 [hep-th]].



\cite{Tasson:2014dfa}
\bibitem{Tasson:2014dfa} 
  J.~D.~Tasson,
  Rept.\ Prog.\ Phys.\  {\bf 77}, 062901 (2014)
  doi:10.1088/0034-4885/77/6/062901
  [arXiv:1403.7785 [hep-ph]].
 
 

%



%




\end{thebibliography}
\end{document}